\documentclass[lettersize,journal]{IEEEtran}
\usepackage{amsmath,amsfonts}
\usepackage{amsthm}
\usepackage{algorithm}
\usepackage{algpseudocode}
\usepackage{amssymb} 
\usepackage{array}
\usepackage[caption=false,font=normalsize,labelfont=sf,textfont=sf]{subfig}
\usepackage{textcomp}
\usepackage{stfloats}
\usepackage{url}
\usepackage{verbatim}
\usepackage{graphicx}
\usepackage{cite}
\usepackage{multirow}
\usepackage{booktabs}
\usepackage{tabularx}
\usepackage{makecell}
\usepackage{hyperref}
\usepackage{xcolor}
\usepackage{enumitem}
\newtheorem{scassumption}{Assumption}
\newtheorem{sclemma}{Lemma}
\newtheorem{sctheorem}{Theorem}

\begin{document}

\title{SplitCom: Communication-efficient Split Federated Fine-tuning of LLMs via Temporal Compression}

\author{Tao Li, Yulin Tang, Yiyang Song, Cong Wu,~\IEEEmembership{Member,~IEEE,} Xihui Liu,~\IEEEmembership{Member,~IEEE,} Pan Li,~\IEEEmembership{Fellow,~IEEE,} Xianhao Chen,~\IEEEmembership{Member,~IEEE}
\thanks{Tao Li, Yiyang Song, Xihui Liu and Xianhao Chen are with the
Department of Electrical and Computer Engineering, University of Hong
Kong, Hong Kong SAR, China. (e-mail: litao@eee.hku.hk; yysong01@connect.hku.hk; xihuiliu@eee.hku.hk; xchen@eee.hku.hk). Yulin Tang is with the Department of Electrical Engineering, California Institute of Technology, California, United States (ytang9@caltech.edu). Cong Wu is with the School of Cyber Science and Engineering, Wuhan University, Wuhan 430072, China. (e-mail: cnacwu@whu.edu.cn). Pan Li is with Hangzhou Dianzi University, China. (email: lipan@hdu.edu.cn). (Corresponding author: Xianhao Chen.)
}}



\maketitle

\begin{abstract}
Federated fine-tuning of on-device large language models (LLMs) mitigates privacy concerns by preventing raw data sharing. However, the intensive computational and memory demands pose significant challenges for resource-constrained edge devices. To overcome these limitations, split federated learning (SFL) emerges as a promising solution that partitions the model into lightweight client-side and compute-intensive server-side sub-models, thus offloading the primary training workload to a powerful server. Nevertheless, the high-dimensional activation exchanges in SFL lead to excessive communication overhead. To overcome this, we propose SplitCom, a communication-efficient SFL framework for LLMs that exploits temporal redundancy in activations across consecutive training epochs. Inspired by video compression, the core innovation of our framework lies in selective activation uploading only when a noticeable deviation from previous epochs occurs. To balance communication efficiency and learning performance, we introduce two adaptive threshold control schemes based on 1) bang-bang control or 2) deep deterministic policy gradient (DDPG)-based reinforcement learning. Moreover, we implement dimensionality reduction techniques to alleviate client-side memory requirements. Furthermore, we extend SplitCom to the U-shape architecture, ensuring the server never accesses clients' labels. Extensive simulations and laboratory experiments demonstrate that SplitCom reduces uplink communication costs by up to 98.8\,\% in its standard configuration and total communication costs by up to 95.8\,\% in its U-shape variant without noticeably compromising model performance.
\end{abstract}


\section{Introduction}
\begin{figure}[t]
    \centering
    \includegraphics[width=0.70\columnwidth]{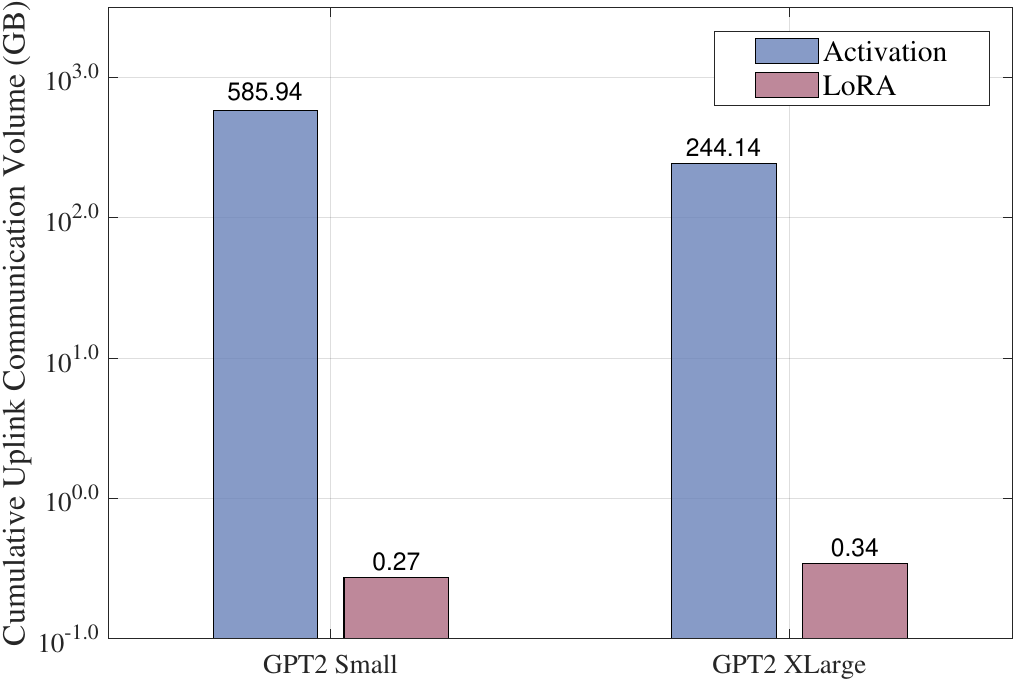}
    \caption{Total uplink communication costs (for 10 clients) for LoRA aggregation and activation uploading until model convergence (GPT2 Small with 50 epochs and GPT2 XLarge with 10 epochs) under split federated LoRA fine-tuning. We partition the E2E dataset into these 10 clients.}
    \label{introduction_1}
    \vspace{-5.5mm}
\end{figure}
In data-silo settings, organizations or individual users are unwilling to share raw data for centralized model training (or fine-tuning) due to privacy concerns and regulatory requirements~\cite{GDPR}. This reluctance severely impedes fine-tuning of pre-trained large language models (LLMs) towards downstream privacy-sensitive applications, such as personal assistants, healthcare, and mental well-being. Federated learning (FL) addresses this issue by enabling clients to locally update models and share only the model parameters with a federated server, thus obviating raw data exchange and safeguarding user privacy~\cite{qu2025mobile}. However, the limited memory and compute budgets of users, particularly edge devices, pose challenges to local fine-tuning of LLMs, even with parameter-efficient fine-tuning (PEFT) such as low-rank adaptation (LoRA)~\cite{hu2021lora}. For example, fine-tuning GPT-2 XLarge (e.g., batch size 4, LoRA rank 24) consumes over 16 GB of GPU memory, which exceeds the capabilities of most edge devices. Against this backdrop, split federated learning (SFL)~\cite{thapa2022splitfed} has been proposed as a compelling alternative capable of overcoming the weaknesses of FL. SFL addresses computing constraints by partitioning deep learning models into client-side and server-side sub-models, thereby offloading the primary training workload to a central server. Consequently, SFL is regarded as a promising distributed paradigm for fine-tuning LLMs under data-silo constraints.

However, applying SFL to fine-tune LLMs still faces a major bottleneck, i.e., \textit{high-dimensional activation communications} at the model split point. Specifically, the frequent transmissions of high-dimensional activations (smashed data) at the cut layer lead to significant communication overhead (shown in Fig.~\ref{introduction_1}). While forward and backward passes both involve communication costs at the cut layer between clients and the server, uplink communication puts an even heavier burden on wireless/mobile devices due to asymmetrical uplink and downlink data rates~\footnote{Since wireless devices are less powerful than base stations, the uplink data rate is often much lower than the downlink data rate in wireless networks, e.g., 30.6 Mbps and 166.8 Mbps for average 5G uplink and downlink, respectively~\cite{wyrzykowski2024hongkong}.} and limited on-device energy. An intuitive approach to mitigate this communication overhead is activation quantization. Nevertheless, since Transformer architectures can generate activations with an extremely wide value range~\cite{sun2024massive, bondarenko2021understanding}, naively quantizing activations can catastrophically degrade the fine-tuning performance for LLMs like GPT variants~\cite{int8}, as validated in Fig.~\ref{introduction_3}. As a result, prior activation-quantization methods designed for traditional split learning (SL) may not work effectively for split fine-tuning of LLMs.

\begin{figure}[t]  
    \centering
    \includegraphics[width=1.0\columnwidth,height=15cm,keepaspectratio]{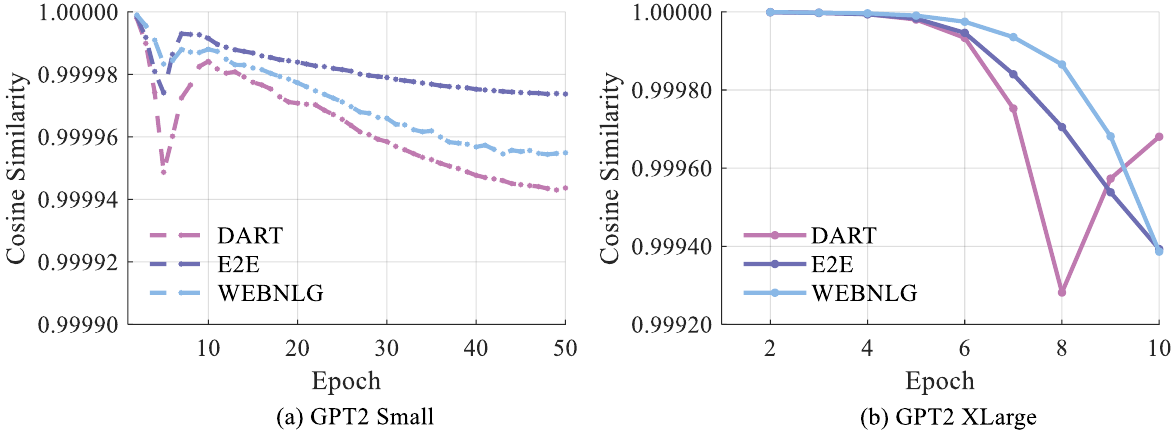}
    \caption{The cosine similarity of activations between the current and previous epoch in split federated fine-tuning for two GPT2 models.}
    \label{introduction_2}
    \vspace{-5.5mm}
\end{figure}
\begin{figure}[t]
    \centering
    \includegraphics[width=0.70\columnwidth]{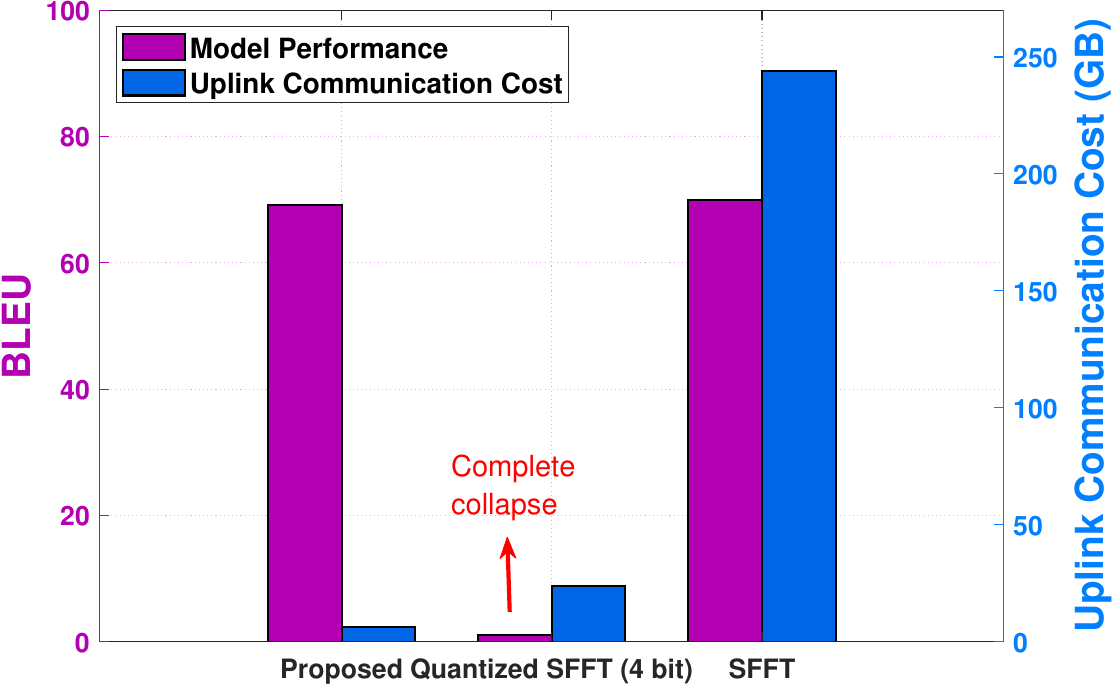}
    \caption{Comparison of the proposed method against traditional SFFT and traditional SFFT with INT4 activation quantization on the GPT-2 XLarge model fine-tuned over the E2E dataset with 10 clients, in terms of uplink communication overhead and model performance. Model performance is measured using BLEU, a standard text generation quality metric, where higher is better.}
    \label{introduction_3}
    \vspace{-6mm}
\end{figure}

To overcome the limitations above, we propose a communication-efficient split federated fine-tuning (SFFT) framework for LLMs that leverages the temporal redundancy in activations across consecutive training epochs. Our approach is grounded in the observation that, given a training sample, since PEFT induces only \textit{tiny changes} to client-side sub-models, the outputs from client-side sub-models evolve very slowly across training epochs. Consequently, activations at the cut layer in SFL are highly correlated between adjacent epochs, as demonstrated in Fig.~\ref{introduction_2}. Motivated by this insight, we introduce \textit{inter-epoch temporal compression}, a module analogous to interframe compression techniques in video compression~\cite{microsoft2021video}~\footnote{In interframe video compression, temporal redundancy between consecutive frames is exploited to transmit only the differences relative to reference frames, significantly reducing data volume.}. Concretely, our framework implements a similarity-aware activation reuse mechanism to boost communication efficiency: Each client maintains a local cache of intermediate activations and only transmits a newly computed activation to the central server if notable deviations from previous epochs are detected. As shown in Fig.~\ref{introduction_3}, by exploiting this temporal redundancy, our scheme substantially reduces communication overhead without noticeably sacrificing model performance, even though 4-bit activation quantization encounters complete training collapse. To safeguard label privacy, we further extend this framework to a \textit{U-shaped architecture} that relocates loss computation to clients, enabling temporal compression on both activations and gradients.

To effectively control the similarity threshold, we employ two adaptive control strategies: 1) the bang-bang control policy, which switches between a conservative mode (low threshold for fewer transmissions) and an aggressive mode (high threshold for more transmissions), simply guided by the dynamics of validation perplexity (PPL)~\cite{radford2019language}. 2) DDPG-driven reinforcement learning (RL), which leverages the deep deterministic policy gradient (DDPG) algorithm~\cite{ddpg} to determine the threshold by balancing communication costs and model performance. The former is rule-based, which is simple to implement, whereas the latter is learning-based. In addition, we implement dimensionality reduction techniques, such as random projection (RP)~\cite{rp}, to minimize the storage requirements for cached intermediate activations and gradients on clients. Our key contributions are summarized as follows:
\begin{itemize}
  \item We introduce SplitCom, a communication-efficient SFFT framework for LLMs with an innovative temporal compression mechanism. We theoretically analyze the convergence bound of SplitCom. We further extend our framework to the U-shape architecture that enables temporal compression on both activations and gradients, which reduces bidirectional communication costs while ensuring that clients' labels are never exposed to the server.
  
  \item We devise two complementary adaptive strategies to determine the similarity threshold based on training dynamics, i.e., a lightweight bang-bang control strategy and a DDPG-based RL strategy.
  
  \item We evaluate SplitCom through simulations and laboratory experiments on three natural language generation (NLG) benchmarks with GPT-2 under IID settings and two natural language understanding (NLU) benchmarks with Qwen-2.5 under non-IID settings. The standard configuration reduces uplink communication by up to \textbf{98.8\,\%}, while the U-shape variant reduces total communication by up to \textbf{95.8\,\%}, both maintaining model performance.
\end{itemize}

The remainder of this paper is organized as follows. Sec.~\ref{related_work} introduces related work. Sec.~\ref{splitcom} presents the design of SplitCom, including the similarity-aware activation reuse mechanism and adaptive threshold control strategies. Sec.~\ref{sec:convergence_analysis} analyzes the impact of stale activation reuse on convergence. Sec.~\ref{u_shape_splitcom} extends SplitCom to bidirectional compression and the U-shape architecture. Sec.~\ref{implementation} introduces the system implementation, followed by performance evaluation in Sec.~\ref{experiments}. Finally, concluding remarks are presented in Sec.~\ref{conclusion}.
\section{Related Work}
\label{related_work}
\textbf{Parameter-efficient fine-tuning.} The staggering size of LLMs, even on-device LLMs, renders their training computationally prohibitive for edge devices like smartphones~\cite{qu2025mobile}. PEFT methods address this by adapting pre-trained LLMs through the introduction and optimization of a small subset of parameters,  while freezing the majority of parameters. Early approaches incorporated adapter modules into Transformer layers; for example, Houlsby \textit{et al.} demonstrated that incorporating and fine-tuning lightweight bottleneck layers, termed adapters (comprising approximately 3.6\% additional parameters), into BERT achieves performance comparable to full fine-tuning~\cite{houlsby2019parameter}. Another parallel approach, LoRA~\cite{hu2021lora}, freezes original model weights and employs trainable low-rank matrices to re-parameterize the model. LoRA is widely recognized as one of the most popular PEFT methods~\cite{flora, wang2024lora}, requiring the tuning of less than 1\% of the parameters needed for full fine-tuning, while achieving comparable performance across diverse downstream tasks without incurring additional inference costs. However, even with PEFT, on-device LLMs remain oversized for on-device fine-tuning on mobile platforms~\cite{li2025mobillm}.

\textbf{Federated fine-tuning of on-device LLMs.} 
In data-silo settings, FL enables decentralized clients to collaboratively train models by exchanging local updates instead of raw data. Given the resource constraints of FL clients, PEFT methods are commonly integrated with FL to facilitate efficient federated fine-tuning. Recently, Kuang et al. introduced FS-LLM~\cite{kuang2024federatedscope}, a toolkit supporting multiple PEFT algorithms, including LoRA, to boost LLM performance in FL while minimizing communication and computation costs. FedIT~\cite{fedit} demonstrated the efficacy of aggregating solely LoRA modules in FL, thereby reducing communication overhead without compromising model performance. Additionally, FLoRA~\cite{flora} addressed aggregation errors in federated LoRA by proposing a noise-free server-side aggregation mechanism. However, these approaches, while effective on high-end GPUs, often underperform on resource-limited clients like mobile devices due to constrained compute and memory resources~\cite{qu2025mobile}.
\begin{figure}[t]       
  \centering
  \includegraphics[width=0.49\textwidth]{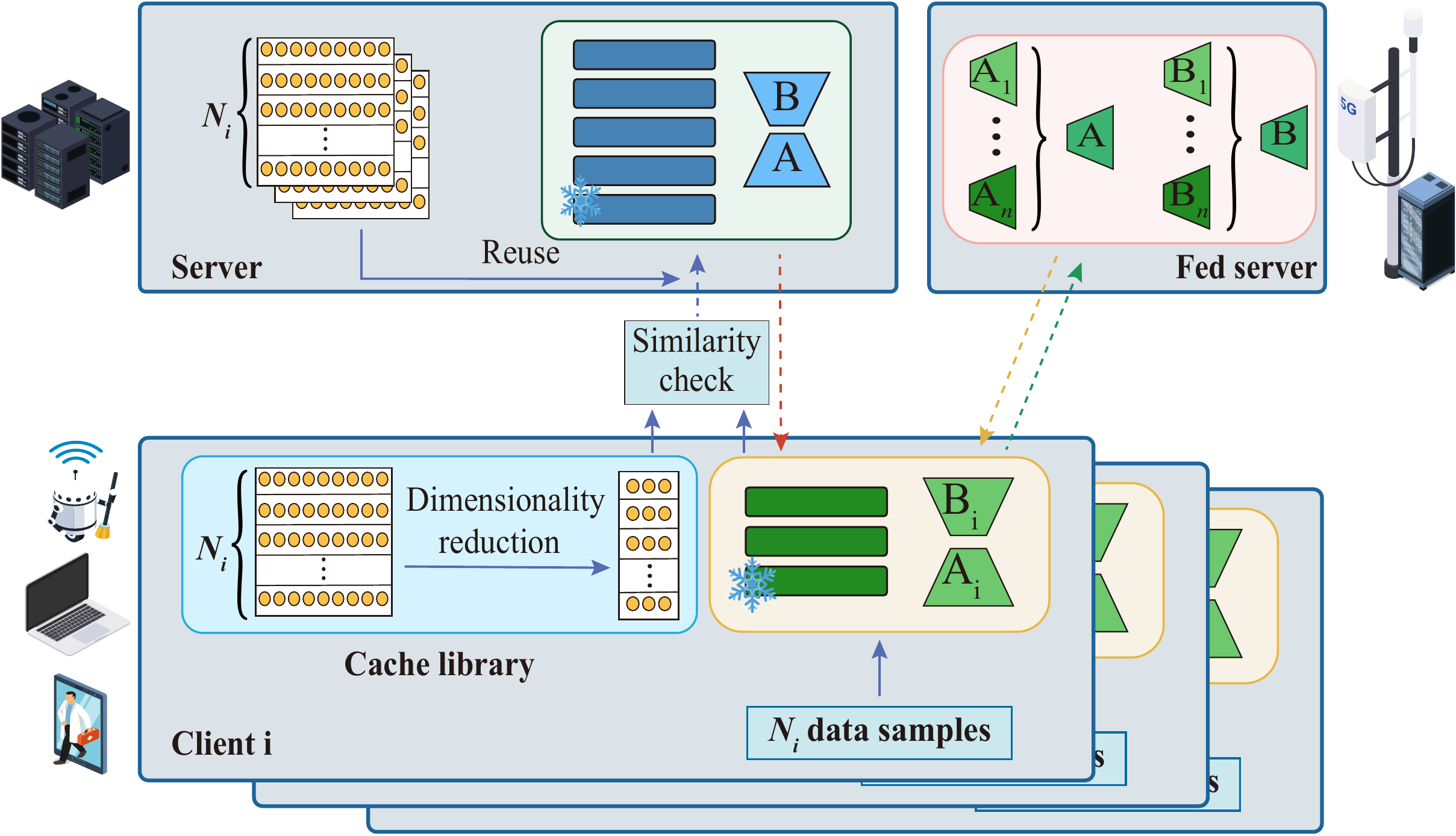}
  \vspace{-4mm}
  \caption{Overview of the SplitCom framework. Clients selectively transmit intermediate activations based on similarity checks, while the server reuses cached activations when transmissions are skipped. By exploiting inter-epoch redundancy, SplitCom reduces communication overhead, and dimensionality reduction further decreases the memory footprint of client-side caches.}
  \label{fig:overview}
  \vspace{-5.5mm}
\end{figure}

\textbf{Split federated fine-tuning on resource-constrained devices.} Federated fine-tuning of LLMs with billions of parameters remains impractical for resource-constrained edge devices. To overcome this, SFL has emerged by combining SL and FL~\cite{thapa2022splitfed}. Lin et al.~\cite{lin2024adaptsfl} analyzed SFL convergence, highlighting cut-layer impacts on performance and optimizing for resource-limited systems. For on-device LLM fine-tuning, combining SFL with PEFT methods like LoRA helps further mitigate resource constraints~\cite{splitlora, chen2025memory}. For instance, SplitLoRA~\cite{splitlora} partitions the model into client-side and server-side sub-models, updating only LoRA adapters to minimize client-side computation overhead~\cite{splitlora}. However, these approaches have yet to address the primary bottleneck of SFL for LLMs, particularly the substantial communication overhead caused by repeated uploads of high-dimensional activations. Furthermore, most existing SFL frameworks~\cite{lin2024adaptsfl, splitlora} require ground-truth labels to reside on the server, inevitably introducing severe privacy leakage risks for sensitive client data. SplitCom targets both limitations, suppressing redundant activation uploads through inter-epoch reuse and supporting a U-shape variant that keeps labels on clients.

\textbf{Communication compression.} While some SL studies quantize cut-layer activations~\cite{oh2025communication}, they are not tailored to LLMs, whose deep architectures and complex activation distributions make training-time compression particularly challenging~\cite{int4}. More advanced methods reduce per-step traffic through activation quantization~\cite{oh2025communication,nips2022}, sparsification~\cite{ijcai2023}, or adaptive compression~\cite{mudvari2024adaptive}, but generally retain communication at every step. Prior stale-information reuse methods target gradients, graph embeddings, or activations from fixed feature extractors~\cite{chen2018lag,peng2022sancus,chawla2024beyond}, rather than cut-layer activations. SplitCom instead exploits the inter-epoch similarity of sample-wise cut-layer activations induced by PEFT and skips their transmission through similarity-gated cache reuse. Moreover, unlike prior adaptive schemes that control aggregation frequency~\cite{luo2024communication} or compression levels~\cite{isik2024adaptive}, SplitCom adapts the activation-reuse threshold using lightweight bang-bang control or DDPG. Its novelty therefore lies in formulating and integrating adaptive temporal activation reuse, rather than in the generic controllers themselves.

\section{Design of SplitCom}
\label{splitcom}
\subsection{Overview}
To mitigate the communication bottleneck in SFL, we propose SplitCom, a communication‑efficient fine‑tuning framework that incorporates temporal compression and parameter‑efficient LoRA updates. The key to SplitCom lies in a similarity-aware activation reuse mechanism, which suppresses transmissions by reusing activations whose changes fall below a threshold (Section~\ref{sec_3.1}). To balance training efficacy and communication cost, we introduce two adaptive threshold tuning strategies: (i) a bang-bang control policy that switches the threshold based on model performance trends and (ii) a DDPG-based reinforcement learning (RL) approach that continuously learns optimal thresholding policies (Section~\ref{sec_3.2}).

SplitCom is based on a standard SFFT framework~\cite{splitlora}. In SFL, clients and the central server exchange intermediate activations at the cut layer during each training step. Formally, let $W = [W_{s,i}; W_{c,i}]$ denote the weights of a pre-trained LLM, partitioned into a server-side sub-model ($W_{s,i}$) and a client-side sub-model ($W_{c,i}$). The cut layers can be different across clients due to their resource heterogeneity~\cite{lin2024adaptsfl}. We introduce trainable LoRA adapters for fine-tuning both parts of the model $W$: $\mathcal{R}_{s, i}$ for the server-side model and $\mathcal{R}_{c,i}$ for the client-side model of each client $i$. During training, only these lightweight LoRA adapters are updated (while $W$ remains frozen), dramatically reducing the number of trainable parameters. Following AdaptSFL~\cite{lin2024adaptsfl}, the server performs
client-wise forward and backward passes sequentially to avoid concurrent computation graphs and reduce peak memory usage. The resulting server-side gradients are accumulated for a single update per training step, matching the client-side update frequency. With heterogeneous cut layers, we decompose each server-side adapter as $\mathcal{R}_{s,i} = [\mathcal{R}_{s}^{\mathrm{com}}; \mathcal{R}_{s,i}^{\mathrm{nc}}]$, where the common part covers server-side layers shared by all clients and the non-common part covers additional layers retained on the server for client $i$. The common part is updated once per training step using gradients accumulated across clients, while each non-common part is updated using only the corresponding client's gradients. With homogeneous cut layers, the non-common part vanishes. After every $M$ steps, clients upload their client-side LoRA adapters $\mathcal{R}_{c,i}$ to the federated server for aggregation~\footnote{We assume homogeneous cut layers across all clients. Our method can be extended to heterogeneous scenarios with different cut layers, and we empirically evaluate this setting in Section~\ref{ablation}.}, after which the averaged client-side adapter is sent to clients to continue training. The overview of SplitCom is illustrated in Fig. \ref{fig:overview}.

The global LoRA adapter is denoted by $\mathcal{R}=[\mathcal{R}_{c};\mathcal{R}_{s}]$, where $\mathcal{R}_{c}$ and $\mathcal{R}_{s}$ are the aggregated client-side and server-side LoRA adapters across all clients, respectively. The objective of split federated LoRA fine-tuning is to find the adapter configuration $\mathcal{R}^*$ that minimizes the following non-convex finite-sum global loss over all $N$ participating devices:
\begin{equation}
    \min_{\mathcal{R}}
    \frac{1}{N} \sum\nolimits_{i=1}^{N}
    f_i(W \mid \mathcal{R}),
    \label{eq:fed_objective}
\end{equation}
where $f_i(W \mid \mathcal{R})$ is the local loss function evaluated on the data of client~$i$ using the frozen global backbone $W$ together with the global LoRA adapter $\mathcal{R}$, i.e., $f_i(W\mid\mathcal{R}) \triangleq
\mathbb{E}_{\xi_i\sim\mathcal{D}_i}
[F_i(W\mid\mathcal{R};\xi_i)]$, where $\xi_i$ captures the training
randomness induced by mini-batch sampling from $\mathcal{D}_i$.
Following the standard stochastic optimization setting~\cite{lin2024adaptsfl, yu2019linear}, the stochastic
gradient is assumed to be an unbiased estimator of the true gradient, i.e.,
$\mathbb{E}_{\xi_i^t\sim\mathcal{D}_i}
[\nabla_{\mathcal{R}}F_i(W\mid\mathcal{R}_i^t;\xi_i^t)
\mid\boldsymbol{\xi}^{[t-1]}] = \nabla_{\mathcal{R}}f_i(W\mid\mathcal{R}_i^t)$,
where
$\boldsymbol{\xi}^{[t-1]}
\triangleq
[\xi_i^\tau]_{i\in\{1,\ldots,N\},\,\tau\in\{0,\ldots,t-1\}}$
denotes all mini-batch sampling randomness up to training step $t-1$.

To solve problem~(\ref{eq:fed_objective}), conventional SFL frameworks for LLM fine-tuning require uploading high-dimensional cut-layer activations at every iteration, resulting in substantial uplink communication overhead. To address this bottleneck, we propose SplitCom, which exploits temporal redundancy in cut-layer activations through similarity-aware activation reuse to reduce uplink communication overhead.

\subsection{Similarity-aware Activation Reuse}
\label{sec_3.1}
Communication between client-side sub-model $W_{c,i}$ and server-side sub-model $W_{s,i}$ is often the major bottleneck in SFL~\cite{thapa2022splitfed}. Unlike earlier efforts that compress every \emph{individual} activation~\cite{liu2022gact, shiranthika2024splitfedzip}, we observe that, under PEFT, e.g., LoRA, the activations of a given sample remain highly correlated across training epochs. This temporal correlation motivates a similarity‑aware activation reuse strategy that curbs redundant transmissions, as detailed below. Furthermore, our approach is orthogonal to bit-packing and other quantization-based compression schemes, as demonstrated in Section~\ref{experiments}.

\textbf{Activation reuse.} Client \(i\) maintains a local cache library during each training epoch, maintaining its own historical activations, whereas the server holds a global cache library comprising entries from all \(N\) clients~\footnote{For scalability, the server can offload activation caches to CPU memory, which can be much larger than GPU memory. For example, an NVIDIA GeForce RTX 4090 provides 24 GB of GPU memory, whereas contemporary servers commonly ship with 128 – 256 GB of system CPU memory, enabling efficient storage of activations.}. For an input sample \(j\) at epoch \(t\), client \(i\) computes the current activation
\(\mathbf{a}_{t,i}^{j}\) and evaluates its cosine similarity against the locally cached counterpart \(\bar{\mathbf{a}}_{c,i}^{j}\) using $\cos\bigl(\mathbf{a}_{t,i}^{j}, \bar{\mathbf{a}}_{c,i}^{j}\bigr)
= \frac{\mathbf{a}_{t,i}^{j} \cdot \bar{\mathbf{a}}_{c,i}^{j}}{\|\mathbf{a}_{t,i}^{j}\|\;\|\bar{\mathbf{a}}_{c,i}^{j}\|}.$ If \(\cos\bigl(\mathbf{a}_{t,i}^{j}, \bar{\mathbf{a}}_{c,i}^{j}\bigr) \ge \theta_t\), client \(i\) skips transmissions and the server reuses \(\bar{\mathbf{a}}_{s, i}^{j}\) for server-side forward and backward passes, where \(\theta_t\) is the similarity threshold balancing communication overhead and model performance. Otherwise, client \(i\) uploads \(\mathbf{a}_{t,i}^{j}\) to the server, and updates its local cache \(\bar{\mathbf{a}}_{c,i}^{j} \gets \mathbf{a}_{t,i}^{j}\). Correspondingly, the server updates \(\bar{\mathbf{a}}_{s,i}^{j} \gets \mathbf{a}_{t,i}^{j}\) to ensure consistency. We hypothesize that when the similarity exceeds \(\theta_t\), uploading $\mathbf{a}_{t,i}^{j}$ or not will not significantly affect gradient updates. When not uploading $\mathbf{a}_{t,i}^{j}$, the server reuses activations from previous rounds, which is equivalent to freezing client-side models during fine-tuning in the current round.
\begin{table}[t]
  \centering
  \renewcommand{\arraystretch}{1.0}
  \caption{Complexity of four feature-similarity metrics.}
  \vspace{-1.5mm}
  \scriptsize 
  \setlength{\tabcolsep}{6pt} 
  \begin{tabular}{lcc}
    \toprule
    \textbf{Metric} &
    \textbf{Computational / Time Complexity} &
    \textbf{Relative Cost vs.\ Cosine} \\
    \midrule
    Cosine  & $O(D)$ &
              Baseline \\
    CKA     & $O(N^{2}D)$ &
              High \\
    CCA     & $O\!\bigl(\min(N,D)^{2}\cdot\max(N,D)\bigr)$ &
              Very high \\
    SVCCA   & PCA: $O(ND^{2})$ $+$ CCA complexity &
              Extremely high \\
    \bottomrule
  \end{tabular}
  \\
  \vspace{0.5mm}
  \footnotesize
  $N$: number of samples; $D$: feature dimensionality.
  \label{tab:complexity_metrics_similarity}
  \vspace{-2mm}
\end{table}
\begin{table}[tbp]
   \centering
   \renewcommand{\arraystretch}{1.05} 
    \caption{Computational complexity of two widely-used dimensionality‑reduction algorithms, where $N$ is the number of samples, $D$ is the original feature dimension, and $K$ is the target dimension ($K \ll D$).}
    \vspace{-1.5mm}
    \scriptsize 
    \setlength{\tabcolsep}{2pt} 
    \begin{tabular}{lcc}
      \toprule
      \textbf{Method} &
      \textbf{Time Complexity} &
      \textbf{Primary Computational Bottleneck} \\
      \midrule
      PCA &
        $\displaystyle O\!\bigl(ND^{2} + D^{3}\bigr)$ &
        Covariance matrix construction and eigen‑decomposition \\
      RP &
        $\displaystyle O(ND K)$ &
        Matrix multiplication for data projection \\
      \bottomrule
    \end{tabular}
     \label{tab:dimred_complexity}
     \vspace{-5mm}
\end{table}

\textbf{Choice of similarity metrics.} Selecting the right similarity metric is essential for efficient activation reuse. As aforementioned, we employ cosine similarity to compare $\mathbf{a}_{t,i}^j$ and cached value $\bar{\mathbf{a}}_{c,i}^{j}$. As shown in \autoref{tab:complexity_metrics_similarity}, we compare cosine similarity~\cite{xia2015learning}, centered kernel alignment (CKA)~\cite{kornblith2019similarity}, canonical correlation analysis (CCA)~\cite{andrew2013deep}, and singular vector canonical correlation analysis (SVCCA)~\cite{raghu2017svcca}. We eventually choose cosine similarity for its computational efficiency, simplicity, and suitability. It requires only simple dot products and two $\ell_2$-norms. In contrast, CCA and SVCCA require the construction and inversion of covariance matrices followed by singular-value decompositions, resulting in prohibitive runtime and memory footprints, while CKA has been shown to exhibit high sensitivity to perturbations in individual samples and incurs several orders of magnitude more computation than cosine similarity~\cite{klabunde2025similarity}. Moreover, CKA, CCA, and SVCCA are invariant to orthogonal or invertible linear transformations, which can mask numerically significant deviations between the current and cached activations. Given the high-dimensional features of LLMs, the compute and memory constraints of client devices, and the need to reliably detect activation deviations, we employ cosine similarity as our metric. Our ablation study in Section~\ref{ablation} empirically confirms that cosine similarity best captures inter-epoch activation redundancy among these four metrics.

\textbf{Cache dimension reduction.} As mentioned earlier, each client maintains a cache library for similarity checks. To mitigate the memory overhead of cached activations on edge devices, we employ dimensionality-reduction techniques, such as RP~\cite{rp} or principal component analysis (PCA)~\cite{abdi2010principal}. As summarized in \autoref{tab:dimred_complexity}, PCA is computationally expensive for high-dimensional data due to the requirements of covariance matrix construction and eigen-decomposition. In contrast, RP relies solely on matrix multiplication and is mathematically proven to preserve pairwise cosine similarity with high probability~\cite{yuan2011efficient}. Therefore, we adopt RP for its computational efficiency and suitability for resource-constrained edge devices. Our extensive experiments in Section~\ref{comparison of pca and rp} further validate this design choice, demonstrating that RP achieves comparable model performance with significantly lower computational overhead on edge devices.
\subsection{Similarity Threshold Control}
\label{sec_3.2}
The similarity-aware activation reuse mechanism significantly reduces communication overhead, but its performance hinges on the selection of the similarity threshold $\theta$. Consider a naive strategy that simply fixes the $\theta$ during the entire training phase, called \textit{fixed control policy}. Such a fixed threshold fails to adapt to the training dynamics, compromising the balance between communication efficiency and model accuracy. Empirically, we observe that early training stages demand frequent communication and a higher $\theta$ to accelerate convergence, while later stages can afford more aggressive reuse with a lower $\theta$ as the model stabilizes. To address this, we propose two adaptive strategies for dynamically adjusting $\theta$: a lightweight bang-bang control policy and a more flexible DDPG-based RL strategy.

\textbf{i) Bang-bang control policy.} To dynamically adjust the similarity threshold~$\theta$ in our SFL framework, we adopt a \emph{bang-bang controller}, a lightweight, rule‑based scheme rooted in classical control theory. The controller switches $\theta$ between a low value $\theta_{\text{low}}$ and a high value $\theta_{\text{high}}$ according to trends of the validation PPL~\cite{radford2019language}, where lower PPL indicates better predictive performance. When PPL consistently decreases, the controller lowers $\theta$ to reduce redundant transmissions. If PPL stops improving or starts to rise, it raises $\theta$ to allow more frequent activation updates. To be specific, the bang-bang controller tracks the validation PPL at the end of every training epoch. Initially, $\theta$ is randomly assigned to either $\theta_{\text{low}}$ or $\theta_{\text{high}}$. As training progresses, the controller evaluates two conditions to decide if an adjustment to $\theta$ is necessary:
\begin{itemize}
    \item \textbf{Switch to \(\theta_{\text{high}}\):} If the current epoch's PPL \(ppl_t\) exceeds the previous epoch's PPL \(ppl_{t-1}\) by a tolerance factor, i.e., \(ppl_t > ppl_{t-1} \cdot (1 + \tau)\), or if a sustained upward trend is detected over a predetermined window size of the previous epochs (e.g., window size = 2 in our setting), the policy sets \(\theta = \theta_{\text{high}}\). This increases activation transmission frequency to maintain model accuracy when convergence falters.
    \item \textbf{Switch to \(\theta_{\text{low}}\):} If \(ppl_t\) decreases, i.e., \(ppl_t < ppl_{t-1}\), for a consecutive number of epochs (e.g., 2), \(\theta\) is set to \(\theta_{\text{low}}\). This mode reduces communication overhead by aggressively exploiting activation reuse.
\end{itemize}
The bang–bang controller is lightweight, as it only compares PPL values, resulting in negligible overhead. This simplicity makes it well-suited to resource‑constrained edge devices. However, its binary action space may miss more subtle trade‑offs between communication cost and model performance in dynamic training environments. To address this, we design the DDPG-based RL strategy to adapt the similarity threshold dynamically.

\textbf{ii) DDPG-based RL strategy.} The DDPG agent consists of an actor network that maps states to actions, and a critic network that estimates the Q-value of state-action pairs. By jointly training these networks, the agent learns to select actions that maximize the Q-value. The two networks are implemented as lightweight architectures, e.g., three-layer fully connected neural networks, in our experiments, making them suitable for edge devices. Unlike the binary adjustments of the bang–bang controller, this strategy learns a continuous policy to dynamically adjust $\theta$. Key components are detailed as follows:

\textbf{(a) State space.} At epoch $t$, the state $s_t$ is represented by a feature vector capturing key training dynamics, including: 1) Training dynamics captured via exponential moving average (EMA)-smoothed similarity scores for robust temporal analysis; 2) Decision context through validation PPL and communication overhead trends, which serve as feedback signals on the action's effectiveness; 3) The normalized training progress $t/T_{\max}$ to guide the action selection process, where $T_{\max}$ is the maximum number of epochs. The compact state provides a comprehensive system status and facilitates the fast convergence of DDPG.

\textbf{(b) Action space.} The action \(a_t\), representing the similarity threshold, is a continuous value in \([0,1]\) and determined by the actor network. A higher threshold increases inter-epoch communication costs. To promote exploration, we incorporate an Ornstein-Uhlenbeck noise process~\cite{lehle2018analyzing, ddpg, hollenstein2022action} with decaying variance.

\textbf{(c) Reward function.} The reward function balances model performance and communication efficiency as $r_t = -\alpha \frac{\ell_t}{\ell_0} - \beta \frac{c_t}{c_0} - P_{\mathrm{zero}} - P_{\mathrm{full}}$, where $\ell_t/\ell_0$ and $c_t/c_0$ are respectively normalized validation loss and communication volume, and $\alpha$, $\beta$ control the model performance-communication efficiency trade-off. The penalty terms $P_{\text{zero}}$ and $P_{\text{full}}$ prevent degenerate communication strategies: zero communication causing model performance degradation and full communication degrading communication efficiency. The agent maximizes the cumulative discounted reward $R = \sum_{t=1}^{T_{\max}} \gamma^{t-1} r_t$ with discount factor $\gamma$.
\begin{algorithm}[t]
\caption{SplitCom training framework}
\label{alg:splitcom}
\setlength{\baselineskip}{0.95\baselineskip}
\begin{algorithmic}[1]
\Require Number of clients $N$; local datasets
$\{\mathcal{D}_i\}_{i=1}^{N}$;
pre-trained weights $W=[W_{s,i};W_{c,i}]$;
local steps per epoch $S$; aggregation interval $M$;
maximum epochs $T_{\max}$; learning rate $\eta$;
threshold policy $\Pi \in \{\mathrm{Fixed},\mathrm{BBC},\mathrm{DDPG}\}$.
\Ensure Trained adapters $\mathcal{R}_{s}^{*}$ and
$\{\mathcal{R}_{c,i}^{*}\}_{i=1}^{N}$.

\State Initialize the server adapter $\mathcal{R}_s$
and all client adapters $\mathcal{R}_{c,i}\gets\mathcal{R}_c^0$, $\forall i\in[N]$; initialize $\theta_1$ according to $\Pi$; freeze $W$ and clear all caches.

\For{each epoch $t=1,\dots,T_{\max}$}
    \For{each local step $m=1,\dots,S$}
        \State Set $k\gets(t-1)S+m$.
        \State Reset accumulated server-side gradients.

        \For{each client $i=1,\dots,N$}
            \State Sample a mini-batch $\mathcal{B}_i$ from $\mathcal{D}_i$ and run \textbf{Algorithm~\ref{alg:activation_reuse}}.
            \State Server performs forward/backward propagation using $\{\bar{\mathbf{a}}_{s,i}^{j}\}$ and accumulates the resulting gradient $\widetilde g_{s,i}^{k}$.
            \State Server sends activation gradients to client $i$.
            \State Update client $i$'s adapter:
                \[
                    \mathcal{R}_{c,i}\gets\mathcal{R}_{c,i}
                    -\eta\widetilde g_{c,i}^{k}.
                \]
        \EndFor

        \State Update the server adapter:
            \[
                \mathcal{R}_{s}\gets\mathcal{R}_{s}
                -\frac{\eta}{N}\sum_{i=1}^{N}
                \widetilde g_{s,i}^{k}.
            \]

        \If{$k \bmod M=0$}
            \State Clients upload
            $\{\mathcal{R}_{c,i}\}_{i=1}^{N}$
            to the federated server.
            \State \textbf{Federated aggregation and broadcast:}
            \[
                \mathcal{R}_{c}\gets
                \frac{1}{N}\sum_{i=1}^{N}\mathcal{R}_{c,i},
                \qquad
                \mathcal{R}_{c,i}\gets\mathcal{R}_{c},
                \quad \forall i\in[N].
            \]
        \EndIf
    \EndFor

    \State Evaluate validation PPL $\mathrm{ppl}_t$
    and communication cost $c_t$.
    \State Update $\theta_{t+1}$ according to $\Pi$.
\EndFor

\State \Return $\mathcal{R}_{s}^{*}$ and
$\{\mathcal{R}_{c,i}^{*}\}_{i=1}^{N}$.
\end{algorithmic}
\end{algorithm}
\begin{algorithm}[t]
\caption{Similarity-aware activation reuse}
\label{alg:activation_reuse}
\setlength{\baselineskip}{0.95\baselineskip}
\begin{algorithmic}[1]
\Require Mini-batch $\mathcal{B}_i$; client-side weights $W_{c,i}$
and adapter $\mathcal{R}_{c,i}$; epoch $t$;
threshold $\theta_t$; client cache
$\{\bar{\mathbf{a}}_{c,i}^{j}\}$ and server cache
$\{\bar{\mathbf{a}}_{s,i}^{j}\}$.
\Ensure Updated caches and server-side activations
$\{\bar{\mathbf{a}}_{s,i}^{j}\}$.

\For{each sample $j$ in $\mathcal{B}_i$}
    \State Compute activation $\mathbf{a}_{t,i}^{j}$
    using $W_{c,i}$ and $\mathcal{R}_{c,i}$.
    \State Obtain $\tilde{\mathbf{a}}_{t,i}^{j}$ via RP.
    \State $s\gets-\infty$.
    \If{$t>1$}
        \State $s\gets\cos(\tilde{\mathbf{a}}_{t,i}^{j},
        \bar{\mathbf{a}}_{c,i}^{j})$.
    \EndIf

    \If{$s<\theta_t$}
        \State Upload $\mathbf{a}_{t,i}^{j}$ to the server.
        \State Update caches:
        $\bar{\mathbf{a}}_{c,i}^{j}\gets
        \tilde{\mathbf{a}}_{t,i}^{j}$,
        $\bar{\mathbf{a}}_{s,i}^{j}\gets
        \mathbf{a}_{t,i}^{j}$.
    \Else
        \State Skip uploading and reuse
        $\bar{\mathbf{a}}_{s,i}^{j}$ on the server.
    \EndIf
\EndFor
\end{algorithmic}
\end{algorithm}

\subsection{The Workflow of SplitCom}
In summary, SplitCom consists of the following steps: 

\textbf{Step 1:} 
SplitCom partitions the pre-trained LLM $W$ into server-side and client-side sub-models using a common cut layer across clients. It initializes the shared server-side adapter $\mathcal{R}_{s}$ and client-side adapters $\{\mathcal{R}_{c,i}\}_{i=1}^{N}$ while keeping $W$ frozen. All caches are initially empty, and $\theta_1$ is set according to policy $\Pi$.

\textbf{Step 2:} 
At each local step, client $i$ computes the cut-layer activation $\mathbf{a}_{t,i}^{j}$ and its RP representation $\tilde{\mathbf{a}}_{t,i}^{j}$ for each sample $j$ in the mini-batch. In the first epoch, all activations are uploaded, with their RP representations cached locally and the original activations cached on the server. In subsequent epochs, an activation is uploaded and both caches are refreshed only if the cosine similarity between its current and cached RP representations is below $\theta_t$; otherwise, the server reuses the cached activation $\bar{\mathbf{a}}_{s,i}^{j}$.

\textbf{Step 3:}
The server processes clients sequentially, performing forward and backward propagation using $\{\bar{\mathbf{a}}_{s,i}^{j}\}$, accumulating server-side gradients, and returning activation gradients to each client for updating $\mathcal{R}_{c,i}$. After processing all $N$ clients, the server updates $\mathcal{R}_{s}$ once using the average of the accumulated gradients.

\textbf{Step 4:} 
Every $M$ local steps, clients upload their adapters to the federated server, which aggregates them via FedAvg~\cite{mcmahan2017communication} and broadcasts the resulting adapter $\mathcal{R}_{c}$ to all clients.

\textbf{Step 5:}
At the end of each epoch, SplitCom evaluates validation PPL $\mathrm{ppl}_t$ and communication cost $c_t$, and sets $\theta_{t+1}$ according to policy $\Pi$; the threshold remains unchanged under the Fixed policy. Training continues for $T_{\max}$ epochs.

The above process is also summarized in Algorithms~\ref{alg:splitcom} and~\ref{alg:activation_reuse}.

\section{Convergence Analysis}
\label{sec:convergence_analysis}
In this section, we mainly analyze the impact of stale activation reuse on the convergence of SplitCom. Suppressing the frozen backbone $W$, let $f_i(\mathcal R)\equiv f_i(W\mid\mathcal R)$ and $F_i(\mathcal R;\xi_i)\equiv F_i(W\mid\mathcal R;\xi_i)$. Recall $\mathcal R_i^{k-1}=[\mathcal R_{c,i}^{k-1};\mathcal R_s^{k-1}]$, $\bar{\mathcal R}_c^k=N^{-1}\sum_i\mathcal R_{c,i}^k$, and $\bar{\mathcal R}^k=[\bar{\mathcal R}_c^k;\mathcal R_s^k]$. The applied gradient is $\widetilde g_{i}^{k} = g_i^k+\delta_i^k$, where $g_i^k=\nabla_{\mathcal R}F_i(\mathcal R_i^{k-1};\xi_i^k)=[g_{c,i}^k;g_{s,i}^k]$ and $\delta_i^k=[\delta_{c,i}^k;\delta_{s,i}^k]$. Each client updates $\mathcal R_{c,i}\gets\mathcal R_{c,i}^{k-1}-\eta(g_{c,i}^k+\delta_{c,i}^k)$, while the server parameters remain fixed during client processing. The server then updates $\mathcal R_s^k=\mathcal R_s^{k-1}-\frac{\eta}{N}\sum_i(g_{s,i}^k+\delta_{s,i}^k)$. Since federated aggregation preserves their average, $\bar{\mathcal R}^{k}=\bar{\mathcal R}^{k-1}-\eta(\bar g^k+\bar\delta^k)$, where $\bar g^k=N^{-1}\sum_i g_i^k$ and $\bar\delta^k=N^{-1}\sum_i\delta_i^k$.

\begin{scassumption}[Smoothness]
\label{ass:sc-smoothness}
Each local loss function $f_i(\mathcal R)$ is differentiable and $L_f$-smooth.
For any $\mathcal R$ and $\mathcal R'$,
\begin{equation}
\label{eq:sc-smoothness}
\left\|\nabla_{\mathcal R}f_i(\mathcal R)-\nabla_{\mathcal R}f_i(\mathcal R')\right\|
\le L_f\|\mathcal R-\mathcal R'\|,\; \forall i.
\end{equation}
\end{scassumption}

\begin{scassumption}[Bounded variance and second moment]
\label{ass:sc-stochastic}
The stochastic gradient is unbiased, and its per-layer variance and second moment are bounded:
\begin{align}
&\mathbb E_{\xi_i\sim\mathcal D_i}
\left\|\nabla_{\mathcal R}F_i(\mathcal R;\xi_i)-\nabla f_i(\mathcal R)\right\|^2
\le \sum_{p=1}^{L}\sigma_p^2,
\;\forall\mathcal R,\ \forall i,
\label{eq:sc-gradient-variance}
\\
&\mathbb E_{\xi_i\sim\mathcal D_i}
\left\|\nabla_{\mathcal{R}_{c}} F_i(\mathcal R;\xi_i)\right\|^2
\le \sum_{p=1}^{L_c}G_p^2,
\;\forall\mathcal R,\ \forall i,
\label{eq:sc-client-gradient-moment}
\end{align}
where $L$ and $L_c$ denote the total number of model layers and the number of client-side layers, respectively, and $\sigma_p^2$ and $G_p^2$ bound the variance and second moment
of the LoRA parameter gradient at layer $p$, respectively.
\end{scassumption}

\begin{scassumption}[Lipschitz regularity]
\label{ass:sc-propagation}
$\nabla_{\mathcal R_s}F_i(\mathcal R_i^{k-1};a)$,
$\nabla_aF_i(\mathcal R_i^{k-1};a)$, and
$\nabla_{\mathcal R_c}F_i(\mathcal R_i^{k-1};G)$
are $L_a^{(\ell)}$-, $H_a^{(\ell)}$-, and $L_G^{(\ell)}$-Lipschitz:
for any $a_1,a_2$ and $g_1,g_2$,
\begin{align}
&\left\|\nabla_{\mathcal R_s}F_i(\mathcal R_i^{k-1};a_1)-\nabla_{\mathcal R_s}F_i(\mathcal R_i^{k-1};a_2)\right\|
\leq L_a^{(\ell)}\|a_1-a_2\|_F, \label{eq:sc-server-gradient-regularity} \\
&\left\|\nabla_aF_i(\mathcal R_i^{k-1};a_1)-\nabla_aF_i(\mathcal R_i^{k-1};a_2)\right\|_F
\leq H_a^{(\ell)}\|a_1-a_2\|_F, \label{eq:sc-activation-gradient-regularity}\\
&\left\|\nabla_{\mathcal R_c}F_i(\mathcal R_i^{k-1};g_1)-\nabla_{\mathcal R_c}F_i(\mathcal R_i^{k-1};g_2)\right\|
\leq L_G^{(\ell)}\|g_1-g_2\|_F, \label{eq:sc-client-backprop-regularity}
\end{align}
with $\ell$ the cut layer.
\end{scassumption}

\begin{sclemma}[Activation reuse error and gradient perturbations]
\label{lem:sc-reuse}
Under Assumption~\ref{ass:sc-propagation} and the
activation-reuse rule in
Algorithm~\ref{alg:activation_reuse},
\begin{equation}
\label{eq:sc-activation-error-control}
\frac{1}{NB}\sum_{i=1}^{N}
\mathbb E\!\left[
\sum_{j\in\mathcal B_i^k}
\|\bar{\mathbf a}_{s,i}^{j,k}
-\mathbf a_{k,i}^{j}\|_F^2
\right]
\le \mathcal E_k,
\end{equation}
\begin{equation}
\label{eq:sc-client-perturbation-bound}
\frac{1}{N}\sum_{i=1}^{N}
\mathbb E\|\delta_{c,i}^{k}\|^2
\le
\bigl(L_G^{(\ell)}H_a^{(\ell)}\bigr)^2
\mathcal E_k,
\end{equation}
\begin{equation}
\label{eq:sc-gradient-perturbation-bound}
\mathbb E\|\bar\delta^{k}\|^2
\le
\frac{1}{N}\sum_{i=1}^{N}
\mathbb E\|\delta_i^{k}\|^2
\le
\Gamma^2\mathcal E_k,
\end{equation}
where $\Gamma=\sqrt{\bigl(L_a^{(\ell)}\bigr)^2 + \bigl(L_G^{(\ell)}H_a^{(\ell)}\bigr)^2}$,
and $\mathcal E_k$ is the per-step activation-reuse error bound defined in (15) and derived in Appendix~B.
\end{sclemma}

\begin{proof}
    See Appendix~B.
\end{proof}

\begin{sclemma}[Client-side LoRA model drift]
\label{lem:sc-drift}
Under Assumption~\ref{ass:sc-stochastic} and
Lemma~\ref{lem:sc-reuse}, the updates satisfy
\begin{equation}
\label{eq:sc-average-client-drift}
\begin{aligned}
&\frac{1}{TN}\sum_{k=1}^{T}\sum_{i=1}^{N}
\mathbb E\left\|
\mathcal R_{c,i}^{k-1}
-\bar{\mathcal R}_c^{k-1}
\right\|^2
\le \eta^2(M-1)^2
\Biggl(
\sqrt{\sum_{p=1}^{L_c}G_p^2}
\\
&\qquad\qquad
+L_G^{(\ell)}H_a^{(\ell)}
\sqrt{\bar{\mathcal E}_T}
\Biggr)^2.
\end{aligned}
\end{equation}
where $M\ge1$ is the client-side aggregation interval,
and $\bar{\mathcal E}_T = \frac{1}{T}\sum_{k=1}^{T}\mathcal E_k$ is the time-averaged activation-reuse error bound.
\end{sclemma}

\begin{proof}
    See Appendix~C.
\end{proof}

\begin{sctheorem}[Main convergence result]
\label{thm:sc-convergence}
Let Assumptions~\ref{ass:sc-smoothness}--\ref{ass:sc-propagation} hold. If $0<\eta\le\frac{1}{L_f}$, then for all $T\ge1$, we have
\begin{equation}
\label{eq:sc-convergence-bound}
\begin{aligned}
&\frac{1}{T}\sum_{k=1}^{T}
\mathbb E\left\|
\nabla f(\bar{\mathcal R}^{k-1})
\right\|^2
\le
\frac{2[f(\bar{\mathcal R}^{0})-f_{\inf}]}{\eta T}
+\frac{L_f\eta}{N}\sum_{p=1}^{L}\sigma_p^2
\\
&\quad+
\Biggl[
L_f\eta(M-1)
\Biggl(
\sqrt{\sum_{p=1}^{L_c}G_p^2}
+L_G^{(\ell)}H_a^{(\ell)}
\sqrt{\bar{\mathcal E}_T}
\Biggr)
+\Gamma\sqrt{\bar{\mathcal E}_T}
\Biggr]^2
\\
&\quad+
2L_f\eta\Gamma
\sqrt{
\frac{\bar{\mathcal E}_T}{N}
\sum_{p=1}^{L}\sigma_p^2
},
\end{aligned}
\end{equation}
where $L$ and $f_{\inf}$ denote the total number of global model layers and the minimum value of problem \eqref{eq:fed_objective}.
\end{sctheorem}

\begin{proof}
    See Appendix~D.
\end{proof}

\textbf{Insights.} The four terms in Theorem~\ref{thm:sc-convergence} respectively capture the initial optimization gap, stochastic gradient noise, the combined effects of client drift and activation reuse, and the interaction between sampling noise and reuse-induced gradient perturbations. Activation reuse affects convergence both directly through
$\Gamma\sqrt{\bar{\mathcal E}_T}$ and indirectly through $L_f\eta(M-1)L_G^{(\ell)}H_a^{(\ell)} \sqrt{\bar{\mathcal E}_T}$, which captures its contribution
to client drift. When current and cached activations align in direction, SplitCom suppresses redundant uploads with limited reuse error. Higher thresholds reduce activation mismatch through more frequent uploads at higher communication cost. This motivates BBC and DDPG to adapt the threshold to training dynamics and balance communication savings against reuse-induced optimization error. A smaller aggregation interval $M$ reduces both the baseline client drift and its amplification of reuse errors, at the cost of more frequent client-side LoRA synchronization.

\section{Extensions to Other SplitCom Variants}
\label{u_shape_splitcom}
The proposed SplitCom framework primarily targets uplink activation transmissions from clients to the server during the forward pass. However, its underlying principles can be naturally extended to downlink gradient transmission and U-shaped SFL architectures. Among these extensions, privacy-preserving U-shaped SFL stands out as a primary deployment scenario for SplitCom, particularly in causal LLM fine-tuning, where server-side loss computation can expose the original training text through shifted-token labels. Its additional communication interfaces further underscore the importance of communication efficiency.

\subsection{Extension to Bidirectional Compression}
In standard SFL, both forward activations (uplink) and backward gradients (downlink) contribute to communication costs. While mitigating uplink traffic is often prioritized due to asymmetric wireless data rates and the limited energy budget of edge devices, suppressing bidirectional communication can further improve overall efficiency. Specifically, the temporal redundancy for activations also applies to gradients: given a training sample, if the model changes very slowly across epochs under PEFT, the corresponding gradients flowing back from the server also exhibit high temporal correlation.

Consequently, the standard SplitCom framework can be extended to compress both uplink activations and downlink gradients by maintaining separate cache libraries on both clients and the server. The server caches activations for potential reuse in forward passes, while clients cache gradients for backward passes. Each transmission direction applies the similarity check independently with its own threshold (e.g., $\theta^\text{up}_t$ for uplink, $\theta^\text{down}_t$ for downlink). For example, when transmitting gradients from server to client $i$ during  the backward pass, for an input sample $j$ at epoch $t$, the server computes the current gradient $z_{t,i}^{j}$, applies RP to obtain its compressed representation $\tilde{z}_{t,i}^{j}$, and evaluates cosine similarity against the compressed cached counterpart  $\bar{\tilde{z}}_{s,i}^{j}$ in its comparison cache using  $\cos\bigl(\tilde{z}_{t,i}^{j}, \bar{\tilde{z}}_{s,i}^{j}\bigr)$.  If $\cos\bigl(\tilde{z}_{t,i}^{j}, \bar{\tilde{z}}_{s,i}^{j}\bigr)  \ge \theta_t^{\text{down}}$, the server skips transmissions and client $i$  reuses the cached gradient $\bar{z}_{c,i}^{j}$ for the frontend backward pass, where $\theta_t^{\text{down}}$ is the similarity threshold. Otherwise, the server transmits the gradient $z_{t,i}^{j}$ to client $i$, and client $i$ updates its cache:  $\bar{z}_{c,i}^{j} \gets z_{t,i}^{j}$. At the same time, the server updates its comparison cache: $\bar{\tilde{z}}_{s,i}^{j} \gets \tilde{z}_{t,i}^{j}$. Moreover, the adaptive threshold management strategies introduced in  Section~\ref{sec_3.2}, specifically bang-bang control and DDPG-based RL,  can be applied to $\theta^\text{down}_t$ to balance downlink communication efficiency and model performance.

\subsection{Extension to U-shape SplitCom}
While the similarity-aware activation reuse mechanism (Section~\ref{sec_3.1}) effectively reduces uplink communication overhead, traditional SFL architectures computing loss on the server expose clients' labels to the server, compromising user privacy.  This is particularly concerning in causal LLM fine-tuning, where labels are shifted input tokens and can thus reveal the original training text. To address these issues, we extend SplitCom to the U-shape architecture~\cite{vepakomma2018split} that relocates loss computation to clients. The model is partitioned into three segments: the client retains the frontend and tail portions\footnote{We define the frontend as the model's initial layers close to the input, and the tail as the final layers proximal to the output.}, while the computationally intensive intermediate layers are offloaded to the server. This configuration allows the loss computation to be performed locally on the client, effectively preserving the privacy of ground-truth labels. However, this U-shape architecture establishes four distinct communication interfaces requiring threshold control: (1) frontend-to-server uplink activations, (2) server-to-tail downlink activations, (3) tail-to-server uplink gradients, and (4) server-to-frontend downlink gradients, which is even more complicated than bidirectional communications. To mitigate the potential communication bottlenecks introduced by these multiple interfaces, we extend our reuse mechanism to operate symmetrically across all four transmission boundaries.

\textbf{Quadri-directional compression.} At each communication session, we apply the similarity-aware reuse mechanism described in Section~\ref{sec_3.1}. Both clients and the server maintain separate cache libraries for activations 
and gradients at each communication session. Specifically, both clients and the server maintain two types of caches: (1) the sender maintains comparison caches, which store RP-compressed representations of activations and gradients to facilitate memory-efficient similarity checks; (2) the receiver maintains reuse caches, which store activations and gradients for actual 
reuse during forward and backward passes when transmission is skipped. This mechanism applies symmetrically to all four communication sessions, each with its own threshold: $\theta_t^{\text{f2s}}$ (frontend-to-server), $\theta_t^{\text{s2t}}$ (server-to-tail), $\theta_t^{\text{t2s}}$  (tail-to-server), and $\theta_t^{\text{s2f}}$ (server-to-frontend). 

The adaptive threshold strategies (BBC and DDPG) introduced in Section~\ref{sec_3.2} can be directly extended to manage thresholds at all four communication sessions in U-shape SplitCom. For the fixed strategy, each communication session is assigned a predetermined constant threshold value that remains unchanged throughout training. For the bang-bang control strategy, each communication session maintains its own pair of low and high thresholds (e.g., $\theta_\text{low}^\text{f2s}$ and $\theta_\text{high}^\text{f2s}$ for the frontend-to-server session), with switching decisions based  on global validation PPL dynamics. For the DDPG-based RL strategy, we deploy  four independent DDPG agents (each consisting of the same lightweight three-layer fully connected  actor-critic networks as described in Section~\ref{sec_3.2}), one for each communication session, where each  agent learns to optimize its respective threshold based on communication efficiency and model performance.

\section{Implementations}
\label{implementation}
In this section, we elaborate on the experimental setup, including datasets and models, baseline methods, hyperparameters, DDPG agent configurations, and implementation details for both simulations and realistic laboratory experiments. 

\textbf{Datasets and models.} We evaluate SplitCom on three representative NLG benchmarks: (1) DART~\cite{dart}, an open-domain data-to-text dataset that maps semantic triples to natural language descriptions; (2) E2E~\cite{e2e}, a restaurant-domain dataset that maps meaning
representations to human-authored texts; and (3) WebNLG~\cite{webnlg}, an RDF-to-text dataset with DBpedia triples covering seen and unseen categories. Each dataset is split into ten clients under IID distribution. Following Hu et al.~\cite{hu2021lora}, we adopt two variants of GPT-2~\cite{radford2019language}: GPT-2 Small, containing 117 million parameters and 12 decoder layers, and GPT-2 XLarge, 
with 1.5 billion parameters and 48 decoder layers. Notably, GPT-2 XLarge is comparable in scale to emerging on-device LLMs such as Google Gemini Nano-1 (1.8B parameters)~\cite{team2023gemini}, making it a representative benchmark for assessing the effectiveness of SplitCom in fine-tuning LLMs on resource-constrained edge devices. We also evaluate the performance of SplitCom on a subset of the GLUE benchmark~\cite{wang2018glue} (MRPC, RTE), to assess its generality on NLU tasks beyond text generation, and on more advanced decoder-only backbones. Specifically, MRPC is a sentence-pair paraphrase-identification task, and RTE is a sentence-pair textual-entailment task; both are formulated as classification problems and evaluated by classification accuracy. For these tasks, we adopt Qwen-2.5-1.5B and Qwen-2.5-3B~\cite{yang2024qwen25}, which represent contemporary on-device LLM families. We partition each dataset across six clients under a Non-IID setting, where Non-IID partitions are generated using a Dirichlet distribution with $\alpha=0.5$~\cite{hsu2019measuring}. To account for training randomness, we repeat each experiment three times and report the mean and standard deviation of both accuracy and communication overhead.

\begin{table}[t]
  \centering
  \small 
  \caption{Client-server model configurations with LoRA adapters in SplitCom.}
  \vspace{-1.5mm}
  \scriptsize
  \setlength{\tabcolsep}{0.125pt} 
    \begin{tabular}{c c c c c c c}
    \toprule
    \multirow{2}{*}{\textbf{Configuration}} &
    \multirow{2}{*}{\textbf{Model}} &
    \multirow{2}{*}{\textbf{Rank}} &
    \multicolumn{2}{c}{\textbf{Client}} &
    \multicolumn{2}{c}{\textbf{Server}} \\
    \cmidrule(lr){4-5} \cmidrule(lr){6-7}
    & & & Total & Trainable & Total & Trainable \\
    \midrule
    \multirow{2}{*}{Standard} & GPT-2 Small & 8 & 60.720M & 0.074M (0.12\%) & 102.610M & 0.221M (0.22\%) \\
    & GPT-2 XLarge & 24 & 174.730M & 0.461M (0.26\%) & 1.470B & 6.912M (0.47\%) \\
    \midrule
    \multirow{2}{*}{U-shape} & GPT-2 Small & 8 & 82.060M & 0.147M (0.18\%) & 42.675M & 0.147M (0.35\%) \\
    & GPT-2 XLarge & 24 & 267.419M & 0.922M (0.34\%) & 1.298B & 6.451M (0.50\%) \\
    \bottomrule
  \end{tabular}
  \label{tab:gpt2_lora_split}
  \vspace{-4mm} 
\end{table}
\begin{figure}[t]
    \centering
    \includegraphics[width=1.0\columnwidth, height=6cm,keepaspectratio]{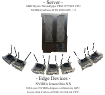}
    \caption{Our SplitCom testbed composed of an RTX\,4090 server and ten Jetson Orin NX clients.}
    \label{testbed}
    \vspace{-4mm}
\end{figure}

\textbf{Baselines.} We evaluate SplitCom in two architectural variants: the standard configuration and the U-shape configuration. For each variant, we compare the following baselines to assess communication efficiency and model performance. (1) \textbf{SplitLoRA~\cite{splitlora}:} the standard SFL with LoRA fine-tuning without activation compression~\cite{thapa2022splitfed}. (2) \textbf{Fixed:} SplitCom with a fixed similarity threshold set to a constant value, without any adaptive control mechanisms, serving as a naive implementation of our temporal compression module. (3) \textbf{BBC:} SplitCom with a lightweight rule-based bang-bang controller (BBC) that adjusts the similarity threshold between $\theta_{\text{low}}$ and $\theta_{\text{high}}$ based on validation PPL dynamics. (4) \textbf{DDPG:} enhanced SplitCom with a DDPG-based RL controller. Our approach is \textit{orthogonal} to prior work on activation and gradient quantization in SFL by introducing a new design dimension. To demonstrate compatibility, we show that our method can be naturally integrated with INT8 quantization~\cite{quantization,li2023model,oh2022communication} to further reduce communication overhead, and append the suffix ``Q'' to each baseline to indicate its INT8-quantized variant\footnote{We evaluate INT8 for activation quantization because INT4 activation quantization is known to cause substantial accuracy degradation in fine-tuning decoder-based LLMs such as GPT variants~\cite{int8}. Moreover, INT4 training requires specialized system support, and unified frameworks for ultra-low precision training remain limited~\cite{int4}.}. In addition, for the NLU tasks under the standard configuration, we further compare against two representative activation/gradient compression baselines originally proposed for distributed and split training: (5) \textbf{AQ-SGD}~\cite{nips2022}, which quantizes cut-layer 
activations based on their temporal variation across training epochs; and (6) \textbf{Rand-Top-$k$}~\cite{ijcai2023}, which sparsifies activations and gradients by randomly retaining the top-$k$ coordinates.

\begin{table*}[t]
\centering
\caption{Performance evaluation of GPT-2 XLarge and GPT-2 Small Models under standard SplitCom on the E2E NLG dataset, where $\uparrow$ means larger is better and $\downarrow$ means lower is better. Communication cost is for uplink communications. Latency is reported in hours. Boldface denotes the best results (highest metric scores and lowest communication/latency costs) excluding SplitLoRA and SplitLoRA\_Q.}
\vspace{-1.5mm}
\small
\setlength{\tabcolsep}{1pt}
\renewcommand{\arraystretch}{0.9}
\begin{tabular}{cccccc| c | c |ccccc| c | c |}
\toprule
& \multicolumn{14}{c}{Dataset (E2E NLG Challenge)} \\
\cmidrule(lr){2-15}
Method & \multicolumn{7}{c}{GPT-2 XLarge} & \multicolumn{7}{c}{GPT-2 Small} \\
\cmidrule(lr){2-8}\cmidrule(lr){9-15}
& BLEU↑ & NIST↑ & MET↑& ROUGE-L↑& CIDEr↑& Comm.↓ & Latency↓ & BLEU↑ & NIST↑ & MET↑& ROUGE-L↑& CIDEr↑& Comm.↓ & Latency↓ \\
\midrule
SplitLoRA & 70.7 & 7.84 & 48.4 & 75.0 & 2.21 & 100\% & 24.68 & 71.7 & 7.99 & 48.0 & 75.0 & 2.28 & 100\% & 55.62 \\
Fixed & 69.4 & 7.86 & \textbf{47.7} & 73.7 & 2.18 & \textbf{10.0\%} & \textbf{7.62} & 69.5 & 7.90 & \textbf{47.4} & \textbf{74.0} & 2.20 & 3.1\% & 11.57 \\
BBC & \textbf{71.0} & \textbf{8.03} & 47.4 & 73.3 & \textbf{2.26} & 11.7\% & 7.94 & \textbf{70.5} & \textbf{8.19} & 46.0 & 72.1 & \textbf{2.25} & \textbf{2.7\%} & \textbf{11.38} \\
DDPG & 69.7 & 7.84 & \textbf{47.7} & \textbf{74.1} & 2.25 & 20.1\% & 9.54 & 69.0 & 8.05 & 45.8 & 70.1 & 2.19 & 2.9\% & 11.51 \\
\midrule
SplitLoRA\_Q & 71.0 & 8.03 & 48.3 & 74.7 & 2.29 & 19.5\% & 9.32 & 55.8 & 5.48 & 37.4 & 64.2 & 1.39 & 6.1\% & 12.63 \\
Fixed\_Q & 67.4 & 7.61 & 47.2 & 73.3 & 2.22 & 3.9\% & 6.46 & 63.2 & 6.20 & 38.8 & 68.1 & 1.52 & \textbf{1.4\%} & \textbf{10.81} \\
BBC\_Q & 68.3 & 7.73 & 47.2 & 73.2 & 2.22 & \textbf{2.6\%} & \textbf{6.22} & 66.0 & 6.34 & 38.0 & 67.3 & 1.58 & \textbf{1.4\%} & \textbf{10.81} \\
DDPG\_Q & \textbf{72.3} & \textbf{8.04} & \textbf{48.5} & \textbf{75.8} & \textbf{2.36} & 10.2\% & 7.66 & \textbf{67.5} & \textbf{8.05} & \textbf{43.9} & \textbf{70.2} & \textbf{2.16} & 3.6\% & 11.09 \\
\bottomrule
\end{tabular}
\vspace{-2mm}
\label{tab:performance-comparison_e2e}
\end{table*}
\begin{table*}[t]
\centering
\caption{Performance evaluation of GPT-2 XLarge and GPT-2 Small Models under standard SplitCom on the DART NLG dataset.}
\vspace{-1.5mm}
\small
\setlength{\tabcolsep}{2pt}
\renewcommand{\arraystretch}{0.9}
\begin{tabular}{cccc| c | c | ccc| c | c |}
\toprule
& \multicolumn{10}{c}{Dataset (DART NLG Challenge)} \\
\cmidrule(lr){2-11}
Method & \multicolumn{5}{c}{GPT-2 XLarge} & \multicolumn{5}{c}{GPT-2 Small} \\
\cmidrule(lr){2-6}\cmidrule(lr){7-11}
& BLEU↑ & MET↑ & TER↓ & Comm.↓ & Latency↓ & BLEU↑ & MET↑ & TER↓ & Comm.↓ & Latency↓ \\
\midrule
SplitLoRA & 35.5 & 0.33 & 0.56 & 100\% & 24.68 & 31.5 & 0.30 & 0.59 & 100\% & 55.62 \\
Fixed & 31.8 & 0.30 & 0.58 & \textbf{10.0\%} & \textbf{7.62} & 29.4 & 0.28 & 0.61 & \textbf{3.2\%} & \textbf{11.61} \\
BBC & 36.2 & 0.33 & \textbf{0.55} & 20.0\% & 9.52 & \textbf{31.3} & \textbf{0.30} & 0.60 & 5.1\% & 12.46 \\
DDPG & \textbf{37.5} & \textbf{0.35} & 0.57 & 20.0\% & 9.53 & 31.1 & \textbf{0.30} & \textbf{0.58} & 3.9\% & 11.93 \\
\midrule
SplitLoRA\_Q & 38.1 & 0.35 & 0.57 & 19.2\% & 9.27 & 19.0 & 0.23 & 0.63 & 6.1\% & 12.59 \\
Fixed\_Q & 32.1 & 0.30 & 0.57 & \textbf{2.0\%} & \textbf{6.09} & 19.5 & \textbf{0.23} & 0.63 & \textbf{1.4\%} & 10.79 \\
BBC\_Q & \textbf{36.5} & \textbf{0.34} & 0.56 & 3.8\% & 6.45 & \textbf{19.8} & \textbf{0.23} & 0.63 & \textbf{1.4\%} & \textbf{10.78} \\
DDPG\_Q & \textbf{36.5} & \textbf{0.34} & \textbf{0.55} & 3.9\% & 6.45 & 19.7 & 0.23 & \textbf{0.62} & 2.1\% & 11.11 \\
\bottomrule
\end{tabular}
\label{tab:performance-comparison_dart}
\vspace{-2mm}
\end{table*}
\begin{table*}[t]
\centering
\caption{Performance evaluation of GPT-2 XLarge and GPT-2 Small Models under standard SplitCom on the WebNLG NLG dataset.}
\vspace{-1.5mm}
\small
\setlength{\tabcolsep}{2pt}
\renewcommand{\arraystretch}{0.9}
\begin{tabular}{cccc| c | c | ccc| c | c |}
\toprule
& \multicolumn{10}{c}{Dataset (WebNLG NLG Challenge)} \\
\cmidrule(lr){2-11}
Method & \multicolumn{5}{c}{GPT-2 XLarge} & \multicolumn{5}{c}{GPT-2 Small} \\
\cmidrule(lr){2-6}\cmidrule(lr){7-11}
& BLEU↑ & MET↑ & TER↓ & Comm.↓ & Latency↓ & BLEU↑ & MET↑ & TER↓ & Comm.↓ & Latency↓ \\
\midrule
SplitLoRA & 70.0 & 0.51 & 0.24 & 100\% & 24.68 & 68.2 & 0.49 & 0.25 & 100\% & 55.62 \\
Fixed & 66.6 & 0.48 & 0.26 & \textbf{10.0\%} & \textbf{7.62} & 49.8 & 0.37 & 0.49 & \textbf{2.5\%} & \textbf{11.29} \\
BBC & \textbf{69.5} & \textbf{0.51} & \textbf{0.25} & 16.1\% & 8.78 & \textbf{68.3} & \textbf{0.49} & \textbf{0.25} & 5.8\% & 12.77 \\
DDPG & 69.3 & 0.50 & \textbf{0.25} & 20.0\% & 9.52 & 66.7 & 0.48 & \textbf{0.25} & 4.0\% & 11.97 \\
\midrule
SplitLoRA\_Q & 70.3 & 0.51 & 0.24 & 19.4\% & 9.31 & 38.5 & 0.32 & 0.45 & 6.0\% & 12.58 \\
Fixed\_Q & 66.7 & 0.48 & 0.26 & \textbf{2.0\%} & \textbf{6.10} & 36.8 & 0.29 & 0.48 & \textbf{1.2\%} & \textbf{10.70} \\
BBC\_Q & 69.2 & 0.50 & 0.25 & 3.2\% & 6.35 & \textbf{37.6} & \textbf{0.32} & \textbf{0.46} & 1.4\% & 10.79 \\
DDPG\_Q & \textbf{71.9} & \textbf{0.51} & \textbf{0.22} & 3.9\% & 6.46 & \textbf{37.6} & \textbf{0.32} & \textbf{0.46} & 1.9\% & 11.02 \\
\bottomrule
\end{tabular}
\label{tab:performance-comparison_webnlg}
\vspace{-2mm}
\end{table*}

Note that we have not included FL baselines because they lead to memory overflow if fine-tuning the entire model on the considered edge devices. For example, fine-tuning GPT-2 XLarge (batch size 4, LoRA rank 24) requires over 16 GB of GPU memory, far exceeding the 8 GB memory capacity of our edge devices.

\textbf{Hyperparameters.}
Following Hu et al.~\cite{hu2021lora}, we apply LoRA to the query ($W_q$) and value ($W_v$) projection matrices in the attention layers. All experiments are conducted with half-precision model parameters for computational efficiency\footnote{Half-precision refers specifically to model parameter precision rather than activations or gradients.}. For the NLG experiments, we have both the standard and U-shape SplitCom configurations.

\textit{Standard SplitCom configuration:}
For the NLG experiments, we deploy 10 clients and use a batch size of 8 for both training and testing. Input sequences are truncated or padded to a maximum length of 512 tokens as necessary. For GPT-2 Small, we set the LoRA rank to $r=8$ and the scaling factor to $\alpha=4$, whereas for GPT-2 XLarge, we use $r=24$ and $\alpha=4$. Each client holds the first 3 decoder layers of the model, while the remaining layers are placed on the server. The detailed model configurations of the clients and the server are illustrated in \autoref{tab:gpt2_lora_split}. To accommodate differences in model scale and convergence behavior, GPT-2 Small and GPT-2 XLarge are trained for 50 and 10 epochs, respectively. The learning rate is $1\times10^{-4}$. The similarity threshold is initialized to 0.98 for all methods. For BBC, the low and high thresholds are set to 0.98 and 0.995, respectively.

\begin{table*}[t]
\centering
\caption{Performance evaluation of GPT-2 XLarge and GPT-2 Small Models under U-shaped SplitCom on the E2E NLG dataset, where $\uparrow$ means larger is better and $\downarrow$ means lower is better. Communication cost is for total (four-way) communication costs. Latency is reported in hours. Boldface denotes the best results (highest metric scores and lowest communication/latency costs) excluding U-shape SplitLoRA and SplitLoRA\_Q.}
\vspace{-1.5mm}
\small
\setlength{\tabcolsep}{1pt}
\renewcommand{\arraystretch}{0.9}
\begin{tabular}{cccccc| c | c |ccccc| c | c |}
\toprule
& \multicolumn{14}{c}{Dataset (E2E NLG Challenge)} \\
\cmidrule(lr){2-15}
Method & \multicolumn{7}{c}{GPT-2 XLarge} & \multicolumn{7}{c}{GPT-2 Small} \\
\cmidrule(lr){2-8}\cmidrule(lr){9-15}
(U-shape) & BLEU↑ & NIST↑ & MET↑& ROUGE-L↑& CIDEr↑& Comm.↓ & Latency↓ & BLEU↑ & NIST↑ & MET↑& ROUGE-L↑& CIDEr↑& Comm.↓ & Latency↓ \\
\midrule
SplitLoRA & 69.2 & 7.87 & 48.1 & 73.8 & 2.25 & 100\% & 100.51 & 69.6 & 7.77 & 47.0 & 72.2 & 2.10 & 100\% & 110.25 \\
Fixed & 67.7 & 7.82 & 46.5 & 71.2 & 2.02 & 35.8\% & 58.95 & 55.8 & 4.77 & 35.6 & 64.2 & 1.27 & 18.6\% & 28.64 \\
BBC & \textbf{70.1} & \textbf{7.87} & \textbf{48.2} & \textbf{74.4} & \textbf{2.26}& \textbf{22.5\%} & \textbf{33.63} & \textbf{62.3} & \textbf{7.39} & \textbf{43.3} & \textbf{69.0} & \textbf{1.64} & \textbf{16.3\%} & \textbf{25.66} \\
DDPG & 68.3& 7.79& 47.2& 73.2& 2.11& 22.8\% & 37.74& 58.2& 5.74& 36.2& 64.4 & 1.37& \textbf{16.3\%} & 25.91\\
\midrule
SplitLoRA\_Q & 71.1 & 7.94 & 49.2 & 75.7 & 2.32 & 11.5\% & 24.92 & 45.0 & 2.27 & 30.5 & 57.7 & 0.86 & 7.2\% & 11.88 \\
Fixed\_Q & 67.7 & 7.68 & 47.0 & 73.6 & 2.14 & 7.3\% & 20.82 & 43.2 & 2.10 & 30.8 & 59.3 & 0.75 & 6.3\% & 10.53 \\
BBC\_Q & 70.0& 7.82& \textbf{48.7} & \textbf{75.6} & 2.28& 7.2\%& 19.30& 46.1& 2.91& 31.2& 58.5& 0.93& \textbf{6.2\%} & \textbf{10.20}\\
DDPG\_Q & \textbf{71.6} & \textbf{7.95} & 48.3& 75.2& \textbf{2.30} & \textbf{4.9\%} & \textbf{14.37} & \textbf{61.6} & \textbf{5.55}& \textbf{35.8} & \textbf{64.8} & \textbf{1.42}& 6.3\%& 11.14\\
\bottomrule
\end{tabular}
\vspace{-2mm}
\label{tab:performance-comparison_e2e-ushape}
\end{table*}
\begin{table*}[t]
\centering
\caption{Performance evaluation of GPT-2 XLarge and GPT-2 Small Models under U-shaped SplitCom on the DART NLG dataset.}
\vspace{-1.5mm}
\small
\setlength{\tabcolsep}{2pt}
\renewcommand{\arraystretch}{0.9}
\begin{tabular}{cccc| c | c | ccc| c | c |}
\toprule
& \multicolumn{10}{c}{Dataset (DART NLG Challenge)} \\
\cmidrule(lr){2-11}
Method & \multicolumn{5}{c}{GPT-2 XLarge} & \multicolumn{5}{c}{GPT-2 Small} \\
\cmidrule(lr){2-6}\cmidrule(lr){7-11}
(U-shape) & BLEU↑ & MET↑ & TER↓ & Comm.↓ & Latency↓ & BLEU↑ & MET↑ & TER↓ & Comm.↓ & Latency↓ \\
\midrule
SplitLoRA & 35.5 & 0.33 & 0.56 & 100\% & 100.55 & 31.3 & 0.30 & 0.58 & 100\% & 110.22 \\
Fixed & 32.4 & 0.32 & 0.59 & 34.2\% & 56.49 & 24.9 & 0.25& 0.71& 17.6\%& 27.41\\
BBC & 35.8& \textbf{0.34} & 0.61 & 21.8\%& 35.95& 24.7 & 0.26 & 0.65 & \textbf{10.2\%} & \textbf{17.16} \\
DDPG & \textbf{36.0} & \textbf{0.34} & \textbf{0.56} & \textbf{20.8\%} & \textbf{34.48} & \textbf{27.2} & \textbf{0.26} & \textbf{0.61} & 12.1\%& 20.30\\
\midrule
SplitLoRA\_Q & 35.1 & 0.33 & 0.56 & 11.3\% & 24.68 & 19.3 & 0.23 & 0.63 & 6.3\% & 10.29 \\
Fixed\_Q & 32.7 & 0.32 & 0.59 & 7.0\% & 19.02 & 17.0 & 0.21 & 0.72 & 5.3\% & 9.63 \\
BBC\_Q & 36.7& \textbf{0.35} & 0.56& 4.8\%& 14.09& 19.1 & \textbf{0.23}& 0.67& \textbf{4.4\%} & \textbf{8.04} \\
DDPG\_Q & \textbf{37.0} & \textbf{0.35} & \textbf{0.55}& \textbf{4.7\%} & \textbf{13.82} & \textbf{19.3} & \textbf{0.23}& \textbf{0.66} & 4.8\%& 8.52\\
\bottomrule
\end{tabular}
\label{tab:performance-comparison_dart-ushape}
\vspace{-2mm}
\end{table*}
\begin{table*}[t]
\centering
\caption{Performance evaluation of GPT-2 XLarge and GPT-2 Small Models under U-shaped SplitCom on the WebNLG NLG dataset.}
\vspace{-1.5mm}
\small
\setlength{\tabcolsep}{2pt}
\renewcommand{\arraystretch}{0.9}
\begin{tabular}{cccc| c | c | ccc| c | c |}
\toprule
& \multicolumn{10}{c}{Dataset (WebNLG NLG Challenge)} \\
\cmidrule(lr){2-11}
Method & \multicolumn{5}{c}{GPT-2 XLarge} & \multicolumn{5}{c}{GPT-2 Small} \\
\cmidrule(lr){2-6}\cmidrule(lr){7-11}
(U-shape) & BLEU↑ & MET↑ & TER↓ & Comm.↓ & Latency↓ & BLEU↑ & MET↑ & TER↓ & Comm.↓ & Latency↓ \\
\midrule
SplitLoRA & 70.6 & 0.51 & 0.23 & 100\% & 100.52 & 68.7 & 0.49 & 0.25 & 100\% & 110.25 \\
Fixed & 68.4 & 0.50 & 0.25 & 34.3\% & 57.89 & 44.6 & 0.40 & 0.56 & 15.4\% & 25.29 \\
BBC & 68.5& 0.50& 0.25& \textbf{20.1\%} & \textbf{34.37} & 50.0 & 0.39 & 0.42 & \textbf{11.7\%} & \textbf{19.78} \\
DDPG & \textbf{69.3}& \textbf{0.50}& \textbf{0.25}& 20.8\%& 34.58& \textbf{52.8} & \textbf{0.42} & \textbf{0.38} & 12.1\% & 20.40 \\
\midrule
SplitLoRA\_Q & 71.4 & 0.51 & 0.22 & 11.1\% & 24.55 & 27.6 & 0.27 & 0.54 & 5.8\% & 9.71 \\
Fixed\_Q & 65.9&  \textbf{0.50} &  0.26&  6.5\%&  15.50& 19.7 & 0.23 & 0.59 & 5.0\% & 9.78 \\
BBC\_Q & 69.2& \textbf{0.50}& \textbf{0.25} & 4.7\%& 14.05& 25.1& \textbf{0.26}& \textbf{0.54}& \textbf{4.2\%} & \textbf{7.76}\\
DDPG\_Q & \textbf{69.7} & \textbf{0.50}& \textbf{0.25} & \textbf{4.4\%} & \textbf{13.59} & \textbf{25.9}& \textbf{0.26}& \textbf{0.54} & 4.7\%& 8.48\\
\bottomrule
\end{tabular}
\label{tab:performance-comparison_webnlg-ushape}
\vspace{-2mm}
\end{table*}

\textit{U-shape SplitCom configuration:}
Each client holds the first 3 decoder layers (frontend) and the last 3 decoder layers (tail), while the intermediate layers are placed on the server. We train GPT-2 Small for 50 epochs with a learning rate of $5\times10^{-5}$ and GPT-2 XLarge for 20 epochs with a learning rate of $2.5\times10^{-5}$. For the Fixed strategy, GPT-2 XLarge uses a uniform threshold of 0.999 across all four communication sessions, whereas GPT-2 Small uses a threshold of 0.99, except for the tail-to-server and server-to-frontend links, for which the threshold is set to 0.999. For BBC, we specify distinct low and high threshold bounds $(\theta_{\text{low}},\theta_{\text{high}})$ for each communication session. Specifically, for GPT-2 XLarge, the bounds are set to $(0.985,0.995)$ for frontend-to-server, $(0.93,0.95)$ for server-to-tail, $(0.925,0.945)$ for tail-to-server, and $(0.925,0.945)$ for server-to-frontend. For GPT-2 Small, the corresponding bounds are $(0.985,0.995)$, $(0.825,0.845)$, $(0.83,0.85)$, and $(0.77,0.79)$, respectively. For DDPG, the initial thresholds for these four communication sessions are set to $\{0.99,0.83,0.84,0.78\}$ for GPT-2 Small and $\{0.985,0.93,0.925,0.925\}$ for GPT-2 XLarge.

For the NLU experiments, we deploy 6 clients and use a batch size of 4 for both training and testing. We set the LoRA rank to $r=16$ and the scaling factor to $\alpha=16$. Each client holds the first 3 decoder layers of the model, while the remaining layers are placed on the server. All models are trained for 10 epochs. The learning rate is $6\times10^{-5}$. The similarity threshold is initialized to 0.98 for all methods. For BBC, the low and high thresholds are set to 0.98 and 0.99, respectively.
\begin{table*}[t]
\centering
\begingroup
\caption{Performance evaluation of Qwen-2.5-3B and Qwen-2.5-1.5B models on the MRPC dataset, where $\uparrow$ means larger is better and $\downarrow$ means lower is better. Communication cost is for uplink communications. Latency is reported in hours. Boldface denotes the best results (highest accuracy and lowest communication/latency costs).}
\vspace{-1.5mm}
\small
\setlength{\tabcolsep}{2pt}
\renewcommand{\arraystretch}{0.9}
\begin{tabular}{cc|c|c|c|c|c}
\toprule
& \multicolumn{6}{c}{Dataset (MRPC NLU Challenge)} \\
\cmidrule(lr){2-7}
Method & \multicolumn{3}{c}{Qwen-2.5-3B} & \multicolumn{3}{c}{Qwen-2.5-1.5B} \\
\cmidrule(lr){2-4}\cmidrule(lr){5-7}
& Accuracy↑ & Comm.↓ & Latency↓ & Accuracy↑ & Comm.↓ & Latency↓ \\
\midrule
SplitLoRA    & $77.9^{(\pm 1.8)}$ & $100^{(\pm 0.0)}\%$          & $1.76^{(\pm 0.00)}$ & $\mathbf{78.3}^{(\pm 0.7)}$ & $100^{(\pm 0.0)}\%$      & $1.29^{(\pm 0.00)}$ \\
BBC          & $77.2^{(\pm 0.5)}$ & $\mathbf{10.4}^{(\pm 0.6)}\%$ & $\mathbf{0.51}^{(\pm 0.00)}$ & $74.5^{(\pm 0.3)}$ & $\mathbf{10.3}^{(\pm 0.1)}\%$ & $\mathbf{0.36}^{(\pm 0.00)}$ \\
DDPG         & $\mathbf{78.4}^{(\pm 1.8)}$ & $11.2^{(\pm 1.7)}\%$ & $0.53^{(\pm 0.02)}$ & $74.5^{(\pm 0.9)}$ & $10.9^{(\pm 0.2)}\%$ & $0.37^{(\pm 0.00)}$ \\
AQ-SGD       & $78.1^{(\pm 1.8)}$ & $32.5^{(\pm 0.0)}\%$  & $0.83^{(\pm 0.00)}$ & $77.0^{(\pm 1.2)}$ & $32.5^{(\pm 0.0)}\%$ & $0.59^{(\pm 0.00)}$ \\
Rand-Top-$k$ & $77.6^{(\pm 2.2)}$ & $55.5^{(\pm 0.0)}\%$  & $0.98^{(\pm 0.00)}$ & $77.4^{(\pm 0.9)}$ & $55.5^{(\pm 0.0)}\%$ & $0.71^{(\pm 0.00)}$ \\
\bottomrule
\end{tabular}
\vspace{-2mm}
\label{tab:performance-comparison_mrpc}
\endgroup
\end{table*}
\begin{table*}[t]
\centering
\begingroup
\caption{Performance evaluation of Qwen-2.5-3B and Qwen-2.5-1.5B models on the RTE dataset.}
\vspace{-1.5mm}
\small
\setlength{\tabcolsep}{2pt}
\renewcommand{\arraystretch}{0.9}
\begin{tabular}{c c|c|c|c|c|c}
\toprule
& \multicolumn{6}{c}{Dataset (RTE NLU Challenge)} \\
\cmidrule(lr){2-7}
Method & \multicolumn{3}{c}{Qwen-2.5-3B} & \multicolumn{3}{c}{Qwen-2.5-1.5B} \\
\cmidrule(lr){2-4}\cmidrule(lr){5-7}
& Accuracy↑ & Comm.↓ & Latency↓ & Accuracy↑ & Comm.↓ & Latency↓ \\
\midrule
SplitLoRA    & $80.8^{(\pm 1.9)}$ & $100^{(\pm 0.0)}\%$          & $1.19^{(\pm 0.00)}$ & $77.9^{(\pm 1.3)}$ & $100^{(\pm 0.0)}\%$          & $0.87^{(\pm 0.00)}$ \\
BBC          & $\mathbf{80.9}^{(\pm 1.1)}$ & $\mathbf{10.0}^{(\pm 0.0)}\%$ & $\mathbf{0.35}^{(\pm 0.00)}$ & $\mathbf{78.0}^{(\pm 1.6)}$ & $\mathbf{10.2}^{(\pm 0.0)}\%$ & $\mathbf{0.25}^{(\pm 0.00)}$ \\
DDPG         & $79.9^{(\pm 1.3)}$ & $11.4^{(\pm 1.5)}\%$ & $0.36^{(\pm 0.01)}$ & $76.9^{(\pm 0.7)}$ & $10.6^{(\pm 0.1)}\%$ & $\mathbf{0.25}^{(\pm 0.00)}$ \\
AQ-SGD       & $79.4^{(\pm 1.6)}$ & $32.5^{(\pm 0.0)}\%$  & $0.56^{(\pm 0.00)}$ & $77.3^{(\pm 0.4)}$ & $32.5^{(\pm 0.0)}\%$ & $0.40^{(\pm 0.00)}$ \\
Rand-Top-$k$ & $80.2^{(\pm 1.3)}$ & $55.5^{(\pm 0.0)}\%$  & $0.66^{(\pm 0.00)}$ & $77.0^{(\pm 1.1)}$ & $55.5^{(\pm 0.0)}\%$ & $0.48^{(\pm 0.00)}$ \\
\bottomrule
\end{tabular}
\vspace{-2mm}
\label{tab:performance-comparison_rte}
\endgroup
\end{table*}

\textbf{DDPG agent configurations.} For the NLG and NLU experiments under the standard configuration, we adopt the same DDPG agent configuration. We implement lightweight DDPG agents with three-layer fully connected networks (hidden dimensions 400-300) for both actor and critic, introducing negligible memory overhead compared to model training. For the standard configuration, the state space comprises EMA-smoothed similarity scores, validation PPL, communication trends, and normalized training progress. The action space outputs a continuous threshold in [0, 1], augmented with Ornstein-Uhlenbeck exploration noise ($\sigma$ = 0.002, decay rate 0.98). The reward function balances normalized validation loss (weight = 2.0) and communication cost (weight = 1.0), with penalty terms to prevent degenerate strategies (zero or full communication). The agent updates at the end of each epoch using mini-batches of size 4 sampled from a replay buffer storing up to 50 experiences for GPT-2 Small and 10 experiences for GPT-2 XLarge. For the U-shape configuration, we deploy four independent DDPG agents, one for each communication session (frontend-to-server, server-to-tail, tail-to-server, server-to-frontend). Each agent receives a 14-dimensional state vector consisting of EMA-smoothed similarities across 10 clients for its specific path, validation PPL trend, communication trend, current threshold, and normalized progress. The action space outputs path-specific continuous thresholds with Ornstein-Uhlenbeck noise ($\sigma$ = 0.005, decay rate 
0.98). The reward function balances normalized forward and backward losses (weight = 1.5) against total communication cost (weight = 2.0), with penalties for degenerate strategies. Training follows the same update protocol as the standard configuration.

\textbf{Testbed setup.} Extensive simulations and realistic laboratory experiments were conducted on the testbed system shown in Fig.~\ref{testbed}. The server is equipped with four NVIDIA RTX~4090 GPUs, each providing \mbox{24\,GB} of GPU memory. The client nodes include ten NVIDIA Jetson Orin~NX devices, each configured with \mbox{8\,GB} of GPU memory. To emulate realistic device-to-server wireless connectivity, we deployed \emph{Google Remote Procedure Call} (\emph{gRPC}) for client-server communication~\cite{grpc2025}. gRPC is a high-performance, open-source RPC framework that runs over HTTP/2 and encodes messages with Protocol Buffers, thereby offering language-agnostic interfaces and low-overhead, bi-directional streaming—properties well aligned with SplitCom's communication requirements. The software stack is implemented using PyTorch v2.4.1 and runs on Python 3.8.10. The grpcio and grpcio-tools packages are standardized at version 1.69.0 to ensure compatibility and reproducibility.
\begin{figure*}[t]
    \centering
    \includegraphics[width=0.85\textwidth]{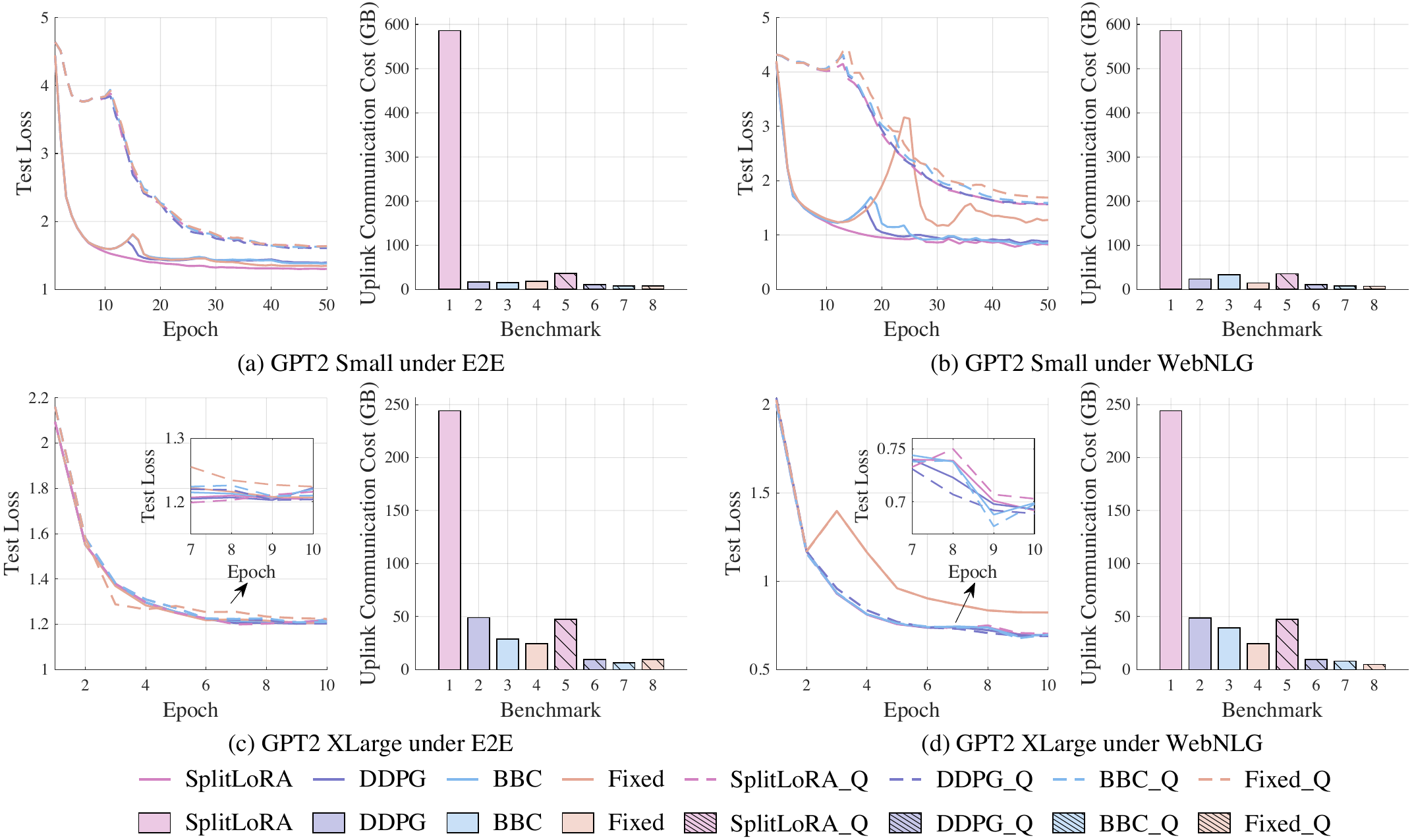}
    \vspace{-1mm}
    \caption{Training dynamics (left axis) and total uplink communication overhead (right-hand bar chart) for two GPT-2 variants under the standard configuration across NLG benchmarks. The total uplink communication overhead accounts for the data transmitted by all clients.}
    \label{fig:training_process_uplink}
    \vspace{-4mm}
\end{figure*}
\section{Experimental Results}
\label{experiments}
In this section, we evaluate the performance of SplitCom from seven aspects: (i) effectiveness of SplitCom in achieving communication reduction while preserving model performance; (ii) generalization to NLU tasks under Non-IID distributions; (iii) integration with INT8 quantization for further efficiency gains; (iv) comparison of adaptive threshold control strategies (BBC and DDPG) to demonstrate the necessity of dynamic adjustment; (v) memory overhead analysis to confirm practical feasibility on resource-limited edge devices; (vi) comparison of dimensionality reduction techniques (PCA vs. RP) to validate our design choice; and (vii) ablation studies on heterogeneous cut layers and similarity metrics.
\begin{figure*}[t]
    \centering
    \includegraphics[width=0.85\linewidth]{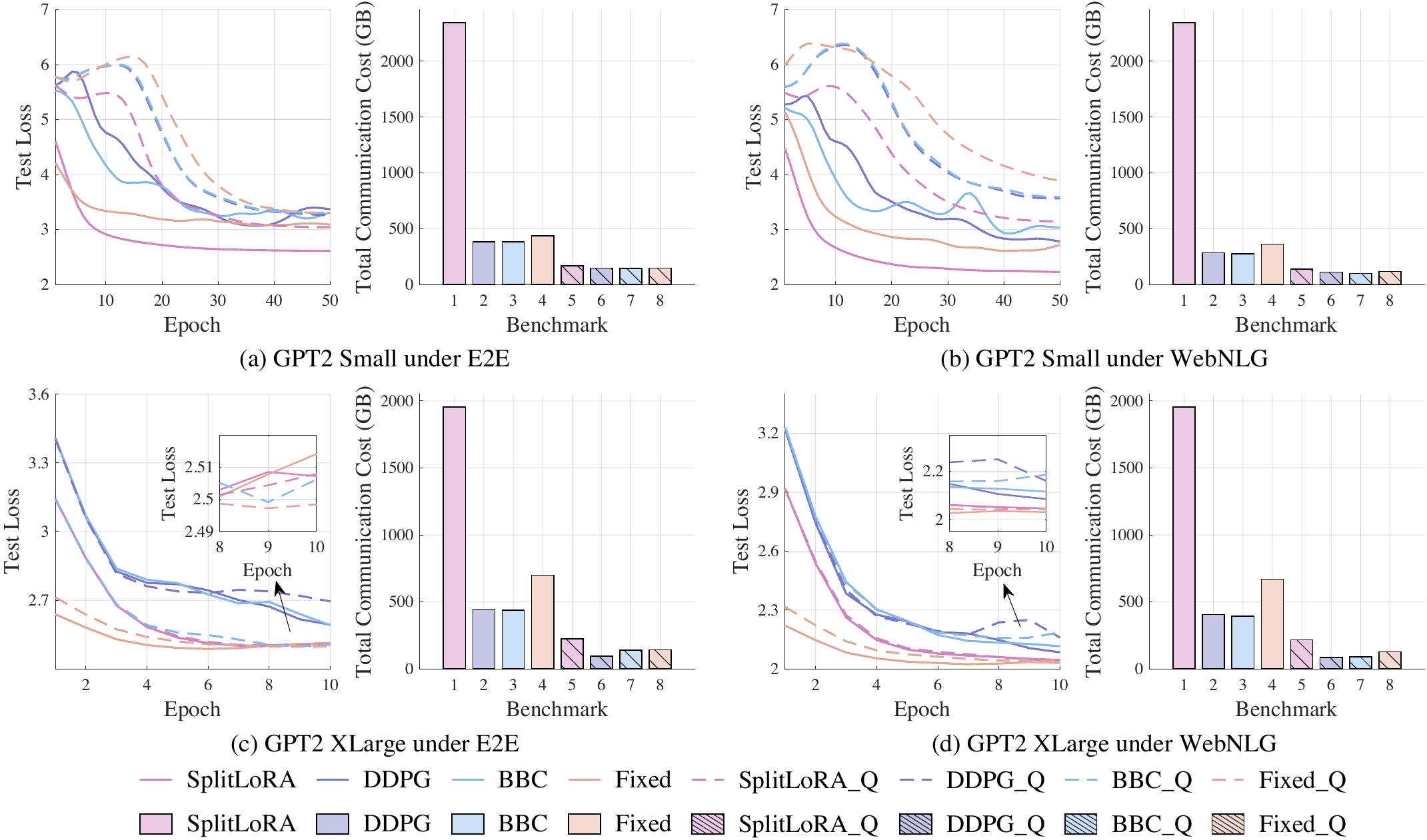}
    \vspace{-1mm}
    \caption{Training dynamics (left axis) and total communication overhead (right-hand bar chart) for two GPT2 variants under the U-shaped configuration across NLG benchmarks. The total communication overhead accounts for the data transmitted by all clients in total.}
    \vspace{-4mm}
    \label{fig:training_process-ushape}
\end{figure*}

\subsection{Effectiveness of SplitCom}
To validate SplitCom, we perform a comprehensive evaluation of NLG accuracy across multiple metrics, as illustrated in~\autoref{tab:performance-comparison_e2e}, ~\autoref{tab:performance-comparison_dart}, \autoref{tab:performance-comparison_webnlg} and Fig.~\ref{fig:training_process_uplink}. The trade-off between BLEU score and uplink communication overhead is shown in Fig.~8 in Appendix~A. Across two GPT-2 variants, SplitCom reduces uplink communication overhead by approximately \textbf{80--97\%}, while preserving language generation performance comparable to the SplitLoRA baseline. Specifically, as illustrated in Fig.~\ref{fig:training_process_uplink}, for GPT-2 XLarge, SplitLoRA requires 244.14 GB, whereas SplitCom utilizes only about 24 GB---achieving a 90\% reduction. For GPT-2 Small, SplitLoRA demands 585.94 GB, in contrast to SplitCom's approximately 16 GB---yielding a 97\% reduction. As shown in Table~\ref{tab:performance-comparison_e2e}, our methods correspondingly decrease latency by up to approximately 80\%. These results support our core insight: LoRA induces minimal updates to client-side sub-models, leading to stable activations across training epochs. This temporal redundancy enables efficient activation reuse with minimal impact on model performance. 

The U-shape architecture enables symmetric temporal compression on both activations and gradients, achieving about \textbf{80--95\%} total communication reduction while strengthening privacy guarantees. As shown in Tables~\ref{tab:performance-comparison_e2e-ushape}, \ref{tab:performance-comparison_dart-ushape}, and \ref{tab:performance-comparison_webnlg-ushape}, and Fig.~\ref{fig:training_process-ushape}, across all three benchmark datasets, U-shape SplitCom with adaptive threshold control (BBC/DDPG) consistently reduces total communication overhead by \textbf{77--80\%} for GPT-2 XLarge and \textbf{84--90\%} for GPT-2 Small compared to the U-shape SplitLoRA baseline, while maintaining comparable or even superior model performance. The trade-off between BLEU score and total communication overhead is shown in Fig.~9 in Appendix~A. Importantly, this architecture eliminates the need for clients to transmit labels to the server, thereby preventing potential information leakage about training objectives---a critical vulnerability in traditional SFL frameworks where server-side loss computation exposes sensitive label information.
\subsection{Effectiveness of SplitCom under Non-IID NLU Settings}
\label{sec:nlu_results}
To examine the generality of SplitCom beyond NLG tasks and GPT-2 backbones, 
we evaluate it on two GLUE NLU benchmarks (MRPC and RTE) using Qwen-2.5-1.5B 
and Qwen-2.5-3B under the Non-IID Dirichlet ($\alpha=0.5$) partitioning 
across six clients, as described in Section~\ref{implementation}. Tables~\ref{tab:performance-comparison_mrpc} 
and~\ref{tab:performance-comparison_rte} report the mean and standard deviation over three runs.

Across both backbones and datasets, SplitCom with adaptive threshold control preserves accuracy comparable to SplitLoRA while reducing uplink communication to only 10.0--11.4\% and latency by up to 72\%. For instance, on RTE with Qwen-2.5-3B (Table~\ref{tab:performance-comparison_rte}), BBC attains an accuracy of 80.9, matching SplitLoRA (80.8), while consuming merely 10.0\% of the uplink traffic and cutting the training latency from 1.19 to 0.35 hours. Compared with the compression baselines, AQ-SGD~\cite{nips2022} and Rand-Top-$k$~\cite{ijcai2023} attain similar accuracy but incur $3\times$ and $5\times$ more communication costs (32.5\% and 55.5\%, respectively). For example, on MRPC with Qwen-2.5-3B (Table~\ref{tab:performance-comparison_mrpc}), DDPG achieves the highest accuracy of 78.4 with only 11.2\% communication overhead, whereas AQ-SGD and Rand-Top-$k$ reach 78.1 and 77.6 at the cost of 32.5\% and 55.5\% communications, respectively.

Notably, the Non-IID partitioning does not undermine the temporal redundancy underpinning SplitCom: each client compares activations against its 
\emph{own} historical cache for the \emph{same} local samples, so inter-client data heterogeneity may not substantially affect intra-client inter-epoch similarity. The consistently small standard deviations (mostly within $\pm 2$ points) further demonstrate robustness against training randomness.

\subsection{Integration with INT8 Quantization} 
As shown in Tables \ref{tab:performance-comparison_e2e}, \ref{tab:performance-comparison_dart} and \ref{tab:performance-comparison_webnlg}, when combined with INT8 quantization, uplink communication overhead is reduced by about \textbf{90--99\%}. For GPT-2 XLarge, we observe that post-quantization activation sparsity remains modest, which has negligible impact on the model's generative capabilities. In contrast, for GPT-2 Small, post-quantization activation sparsity becomes extremely high, significantly degrading the model's generative performance. This set of experiments demonstrates that SplitCom can be seamlessly integrated with standard approaches for reducing communication overhead in SL, such as activation quantization, thereby achieving further improvements in communication efficiency. As presented in Tables \ref{tab:performance-comparison_e2e-ushape}, \ref{tab:performance-comparison_dart-ushape} and \ref{tab:performance-comparison_webnlg-ushape}, U-shape SplitCom can also integrate seamlessly with INT8 quantization, reducing total communication overhead by approximately \textbf{93--96\%}. A key observation across models is that backward gradient sparsity remains inherently high and stable with or without quantization, suggesting that performance variations are primarily driven by changes in activation sparsity. Specifically, for GPT-2 XLarge, gradient sparsity remains largely stable at a high magnitude post-quantization; in contrast, forward activation sparsity undergoes a substantial transition from dense to moderately sparse. This specific sparsity pattern has negligible impact on generative capabilities; for instance, as presented in Table~\ref{tab:performance-comparison_e2e-ushape}, DDPG\_Q achieves a BLEU score of 71.6 on E2E while reducing communication to only 4.9\% of the U-shape SplitLoRA. In contrast, for GPT-2 Small, while gradient sparsity exhibits small variation at high magnitude, the forward activation sparsity transitions from a dense state to a heavily sparse regime. It is this substantial increase in activation sparsity that significantly degrades the model's generative performance. These results demonstrate that while U-shape SplitCom naturally exhibits asymmetric sparsity (higher gradient sparsity), larger models possess greater robustness to the activation sparsity induced by INT8 quantization compared to their smaller counterparts.

\begin{table}[t]
  \centering
  \renewcommand{\arraystretch}{1}
    \caption{Cache costs (in CPU memory) for clients and the server to support activation and gradient reuse. Standard configuration caches activations only, while U-shape configuration caches both activations and gradients. In our experiments, the client CPU memory and server CPU memory are 8GB and 256GB, respectively. The percentages in brackets represent the ratios of cache costs relative to the total CPU memory space.}
    \vspace{-1.5mm}
   \label{tab:gpt2_memory}
    \scriptsize
    \setlength{\tabcolsep}{0.125pt} 
    \begin{tabular}{c c c c}
      \toprule
      \textbf{Configuration} &
      \textbf{Model} & 
      \textbf{Client Cache Library (GB)} &
      \textbf{Server Cache Library (GB)}\\
      \midrule
      \multirow{2}{*}{Standard} & GPT-2 Small & 0.20 (2.5\%) & 11.72 (4.6\%) \\
      & GPT-2 XLarge & 0.40 (5.0\%) & 24.41 (9.5\%) \\
      \midrule
      \multirow{2}{*}{U-shape} & GPT-2 Small & 2.73 (34.1\%) & 27.34 (10.7\%) \\
      & GPT-2 XLarge & 5.27 (65.9\%) & 52.73 (20.6\%) \\
      \bottomrule
    \end{tabular}
   \vspace{-4mm}
\end{table}

\subsection{Comparison of BBC and DDPG}
Our work devises two adaptive threshold controllers, i.e., bang-bang control and DDPG-based policy. In general, BBC and DDPG strategies perform similarly well. For instance, as shown in Table~\ref{tab:performance-comparison_dart}, on the DART dataset using GPT-2 XLarge, DDPG shows modest improvements over BBC in BLEU (37.5 vs. 36.2) and METEOR (0.35 vs. 0.33) scores, but with slightly worse TER (0.57 vs. 0.55) and comparable communication overhead and latency (9.53 hours vs. 9.52 hours). In contrast, as shown in Tables~\ref{tab:performance-comparison_e2e} and \ref{tab:performance-comparison_webnlg}, on the E2E and WebNLG datasets, BBC outperforms DDPG in certain metrics such as BLEU (e.g., 71.0 vs. 69.7 on E2E with GPT-2 XLarge).

On the other hand, we observe from Tables~\ref{tab:performance-comparison_e2e} and \ref{tab:performance-comparison_webnlg} that the Fixed method, a naive implementation of our temporal module without adaptive threshold control, still performs remarkably well. However, due to the lack of adaptability, it can exhibit a non-trivial performance gap compared with BBC and DDPG and exhibit relatively severe performance fluctuations in the training process, as shown in Fig.~\ref{fig:training_process_uplink}.

For U-shape SplitCom, the BBC and DDPG strategies exhibit comparable overall model effectiveness, yielding similar trends in communication reduction. Specifically, in Table~\ref{tab:performance-comparison_e2e-ushape}, on the E2E dataset with GPT-2 XLarge, BBC attains a BLEU score of 70.1, outperforming DDPG (68.3) while maintaining nearly identical communication overhead. Conversely, DDPG demonstrates modest advantages on other datasets; for instance, on DART and WebNLG, DDPG slightly surpasses BBC in BLEU scores (36.0 vs. 35.8 and 69.3 vs. 68.5, respectively) while preserving equivalent communication efficiency.

In contrast, the Fixed method, limited by its static nature, yields suboptimal results in both model generative performance and communication efficiency. For instance, in Table~\ref{tab:performance-comparison_e2e-ushape}, on E2E with GPT-2 XLarge, the Fixed approach not only degrades generation quality (BLEU 67.7) but also incurs a significantly higher communication overhead of 35.8\%, compared to 22.5\% for BBC and 22.8\% for DDPG. This consistent inefficiency, characterized by higher communication costs yielding inferior model generative performance, underscores the critical necessity of adaptive threshold control for effectively balancing the bidirectional communication-performance trade-off in U-shape architectures.

For these two threshold control schemes, each method can outperform the other. In practice, bang-bang control can be employed when adapting to entirely new scenarios (i.e., fine-tuning a new LLM under new datasets), whereas DDPG can be used if SplitCom is implemented in similar circumstances where the reinforcement learning agent has learned from past training experience. We encourage the community to treat these two schemes not as final answers but as reference points for subsequent work on communication-efficient SFL with the innovative temporal compression module.

\subsection{Cache Costs on Clients and Server}
To reduce client-side memory overhead, we apply dimensionality reduction (i.e., RP) to compress the activation-cache library. For example, activation tensors are reduced from \(512 \times 1600\) to \(512 \times 256\) for GPT-2 XLarge. As shown in \autoref{tab:gpt2_memory}, the resulting client-side cache requires only 0.2\,GB for GPT-2 Small and 0.4\,GB for GPT-2 XLarge. On the server side, original (uncompressed) activations must be stored for reuse, consuming 11.72\,GB and 24.41\,GB for GPT-2 Small and GPT-2 XLarge, respectively. Given typical CPU memory capacities (say, 8\,GB on clients and 256\,GB on servers in our setup), these storage requirements are well within practical resource limitations. In the case of U-shape SplitCom, cache requirements increase because both activations and gradients must be stored bilaterally to enable bidirectional temporal compression. Specifically, client-side usage rises to 2.73\,GB for GPT-2 Small and 5.27\,GB for GPT-2 XLarge, while server-side consumption increases to 27.34\,GB and 52.73\,GB, respectively. Despite this increased overhead compared to the unidirectional setup, the storage requirements remain within the aforementioned practical resource limitations (8\,GB/256\,GB) for edge deployments of on-device LLMs.
\begin{table}[t]
  \raggedright
  \centering
  \renewcommand{\arraystretch}{1.25}
  \caption{Comparison of PCA and RP schemes on GPT2 XLarge across NLG benchmarks. Each cell reports PPL with uplink communication overhead relative to SplitLoRA. Smaller PPL indicates superior model performance, and a smaller communication-overhead ratio denotes greater communication efficiency.}
  \vspace{-1.5mm}
  \scriptsize
  \setlength{\tabcolsep}{2pt} 
  \begin{tabular}{lcc|cc}
    \toprule
    \multirow{2}{*}{\textbf{Dataset (Dim.\ Red.)}} &
      \multicolumn{2}{c}{\textbf{Unquantized}} &
      \multicolumn{2}{c}{\textbf{Quantized}} \\ 
    \cmidrule(lr){2-3}\cmidrule(lr){4-5}
      & \textbf{BBC} & \textbf{DDPG}
      & \textbf{BBC\_Q} & \textbf{DDPG\_Q} \\
    \midrule
    DART (PCA)
      & \makecell[c]{5.052 / 19.9\%} & \makecell[c]{5.169 / 29.2\%} 
      & \makecell[c]{5.084 / 5.0\%}  & \makecell[c]{5.633 / 7.2\%}  \\
    DART (RP)
      & \makecell[c]{\textbf{4.613} / \textbf{10.0\%}} & \makecell[c]{\textbf{4.613} / \textbf{10.8\%}} 
      & \makecell[c]{\textbf{4.654} / \textbf{2.5\%}}  & \makecell[c]{\textbf{4.654} / \textbf{2.5\%}}  \\
    E2E (PCA)
      & \makecell[c]{\textbf{4.310} / 25.2\%} & \makecell[c]{\textbf{4.295} / 29.5\%} 
      & \makecell[c]{\textbf{4.360} / 7.3\%}  & \makecell[c]{\textbf{4.283} / 6.5\%}  \\
    E2E (RP)
      & \makecell[c]{4.368 / \textbf{20.0\%}} & \makecell[c]{4.312 / \textbf{20.3\%}} 
      & \makecell[c]{4.419 / \textbf{5.0\%}}  & \makecell[c]{4.361 / \textbf{6.3\%}}  \\
    \bottomrule
\end{tabular}
\vspace{-4mm}
\label{pca_random_projection_gpt2xlarge}
\end{table}

\subsection{Comparison of PCA and RP}
\label{comparison of pca and rp}
To address the memory constraints of resource-limited edge devices, we apply a dimensionality reduction process to the cache of activations in Section \ref{sec_3.1}. We compared two techniques—PCA and RP—based on system performance and computational overhead, ultimately selecting RP. Experimental results, detailed in Table~\ref{pca_random_projection_gpt2xlarge}, show that RP achieves comparable or even superior validation PPL to PCA while reducing uplink communication overhead by as much as 80–90\%. Additionally, as illustrated in Section~\ref{sec_3.1}, RP incurs significantly lower computational overhead than PCA, making it more suitable for real-time, latency-critical scenarios~\cite{wojnowicz2016projecting}. 

PCA may also overfit idiosyncrasies of the training data because its projection directions depend heavily on the observed covariance structure. In contrast, the inherent stochasticity of RP acts as a form of regularization, mitigating overfitting while provably preserving pairwise cosine similarity with high probability — a property rooted in its interpretation as a locality-sensitive hashing scheme for cosine distance~\cite{yuan2011efficient}.

Empirically, Table~\ref{pca_random_projection_gpt2xlarge} and Table~XVI in Appendix~A show that RP delivers model performance and uplink communication savings comparable to, or better than, PCA across multiple datasets and two GPT-2 variants. Overall, these theoretical and empirical advantages motivate our exclusive use of RP for dimensionality reduction in SplitCom.
\begingroup
\subsection{Ablation Study}
\label{ablation}
\endgroup
In this subsection, we conduct ablation studies on two key aspects of SplitCom: the choice of cosine similarity for activation reuse and the impact of cut layer heterogeneity across clients. All experiments are performed on the NLU tasks under the standard configuration.

\textbf{Choice of similarity metrics.} Section~\ref{sec_3.1} demonstrates cosine similarity is most computationally efficient; here we empirically verify that it also best captures activation redundancy. Table~\ref{tab:similarity-metrics-comparison} compares cosine, CKA, CCA, and SVCCA under the BBC policy on Qwen-2.5-3B. Cosine similarity achieves the highest accuracy on both tasks (77.7 on MRPC and 82.0 on RTE) at the lowest communication cost (10.0\%), whereas CKA, CCA, and SVCCA yield 4--6 points lower accuracy on MRPC and higher communication overhead on RTE (up to 16.7\%). These results indicate that, beyond its $O(D)$ efficiency, cosine similarity is also the most reliable indicator of inter-epoch activation redundancy in our setting.

\begin{table}[t]
\centering
\begingroup
\caption{Performance comparison of different similarity metrics on the BBC baseline. Each cell reports accuracy with uplink communication overhead relative to SplitLoRA.}
\vspace{-2mm}
\label{tab:similarity-metrics-comparison}
\renewcommand{\arraystretch}{1.25} 
\scriptsize
\setlength{\tabcolsep}{6pt} 
\begin{tabular}{cccc}
\toprule
Model & Metric & MRPC & RTE  \\
\midrule
\multirow{4}{*}{\shortstack{Qwen-2.5\\(3B)}}
& Cosine     & \textbf{77.7}/\textbf{10.0\%} & \textbf{82.0}/\textbf{10.0\%}   \\
& CKA        & 71.6/10.1\% & 80.5/16.7\%  \\
& CCA        & 73.0/10.1\% & 80.5/16.7\%  \\
& SVCCA      & 73.0/10.1\% & 80.9/14.3\%  \\
\bottomrule
\end{tabular}
\vspace{-2mm}
\endgroup
\end{table}
\begin{table}[t]
\centering
\begingroup
\caption{Performance comparison of BBC and DDPG under homogeneous and heterogeneous cut layer settings.}
\label{tab:bbc_ddpg_homo_hetero}
\vspace{-2mm}
\renewcommand{\arraystretch}{1.25}
\scriptsize
\setlength{\tabcolsep}{2pt} 
\begin{tabular}{ccc|cc|c}
\toprule
\multirow{2}{*}{\textbf{Model}} & \multirow{2}{*}{\textbf{Task}} & \multicolumn{2}{c}{\textbf{BBC}} & \multicolumn{2}{c}{\textbf{DDPG}} \\
\cmidrule(lr){3-4}\cmidrule(lr){5-6}
& & Homogeneous & Heterogeneous & Homogeneous & Heterogeneous \\
\midrule
\multirow{2}{*}{\shortstack{Qwen-2.5\\(1.5B)}}
& MRPC & 74.8/10.2\% & 75.0/10.6\% & 75.0/10.7\% & 75.0/12.2\% \\
& RTE  & 79.1/10.2\% & 76.2/10.2\% & 76.9/10.5\% & 80.9/11.9\% \\
\midrule
\multirow{2}{*}{\shortstack{Qwen-2.5\\(3B)}}
& MRPC & 77.7/10.0\% & 80.9/17.4\% & 80.4/13.2\% & 77.2/15.0\% \\
& RTE  & 82.0/10.0\% & 81.2/10.0\% & 81.2/13.1\% & 81.2/16.7\% \\
\bottomrule
\end{tabular}
\vspace{-4mm}
\endgroup
\end{table}

\textbf{Impact of heterogeneous cut layers.} To assess SplitCom under client resource heterogeneity, we compare homogeneous cut layers (all six clients split at layer 3) against a heterogeneous setting ($\{3,3,2,2,3,3\}$), where the heterogeneous split-point scheme follows AdaptSFL~\cite{lin2024adaptsfl}. As shown in Table~\ref{tab:bbc_ddpg_homo_hetero}, heterogeneous cut layers incur comparable or higher communication costs, with accuracy effects varying across models, tasks, and controllers. A possible explanation is that client-specific representation depths produce diverse activation-similarity distributions, making a shared threshold policy less uniformly suited to all clients. Maintaining accuracy may therefore require more frequent activation transmissions, which increase communication but reduce approximation errors from cached-activation reuse, potentially improving accuracy in some configurations. For example, BBC improves accuracy from 77.7\% to 80.9\% on Qwen-2.5-3B under MRPC, while DDPG increases it from 76.9\% to 80.9\% on Qwen-2.5-1.5B under RTE. Despite this overhead, both controllers retain substantial savings, requiring only 10.0\%--17.4\% of SplitLoRA's communication cost. Moreover, since similarity checks and activation reuse operate independently on each client's cache, SplitCom accommodates heterogeneous cut layers without modifying its temporal reuse mechanism.
\vspace{2mm}
\section{Conclusion}
\label{conclusion}
In this work, we identify that existing SFL frameworks suffer from communication bottlenecks due to frequent client–server exchanges. To address this, we introduce SplitCom, a communication‐efficient SFFT framework for LLMs that exploits temporal redundancy in activation uploading. By integrating similarity‐aware activation reuse and bang–bang control or DDPG-based RL, SplitCom dramatically reduces uplink communication overhead with negligible loss in model performance. To simultaneously achieve comprehensive communication efficiency and strengthen privacy guarantees, we further extend SplitCom to a U-shaped architecture that relocates loss computation to clients, enabling symmetric temporal compression on both activations and gradients while ensuring that clients' labels are never exposed to the server. Our extensive evaluations across NLG and NLU tasks with GPT-2 and Qwen-2.5 backbones on both high-end GPU servers and Jetson Orin NX edge devices demonstrate that SplitCom, in both its standard and privacy-enhanced U-shape configurations, makes on-device LLM application practical under stringent computational and communication constraints.

Future directions will focus on algorithmic synergy. To address the performance variability observed in smaller models like GPT-2 Small under INT8 quantization, we propose to explore a co-design approach. This involves integrating adaptive quantization (such as dynamic bit-allocation) with temporal compression to achieve a robust balance between compression rate and generative performance.

\bibliographystyle{IEEEtran}
\bibliography{reference}

\end{document}